\documentclass[twocolumn]{aastex631}

\usepackage{gensymb}
\usepackage{natbib}      
\usepackage{graphicx}
\usepackage{amsmath}
\usepackage{times}
\usepackage{amssymb}
\usepackage{bm}
\usepackage{fancyhdr}
\usepackage{epsfig}
\usepackage{float}
\usepackage{url}
\usepackage{array}
\usepackage{verbatim}     
\usepackage{calc,layouts}
\usepackage{boldline}
\usepackage{tabu}
\usepackage{rotating}     
\usepackage{xcolor,colortbl}
\usepackage{multirow}

\hypersetup{linkcolor=red,citecolor=blue,filecolor=cyan,urlcolor=magenta}

\newcommand{\p}[1]{{\color{magenta}{#1}}}
\newcommand{\q}[1]{{\color{red}{#1}}}

\begin{document}
%

\title{Deflection of Coronal Mass Ejections in Unipolar Ambient Magnetic Fields}

\correspondingauthor{Michal Ben-Nun}
\email{mbennun@predsci.com}

\shorttitle{CME Deflection}
\shortauthors{Ben-Nun et al.}

\author[0000-0002-9164-0008]{Michal~Ben-Nun}
\affiliation{Predictive Science Inc., 9990 Mesa Rim Road, Suite 170, San Diego, CA 92121, USA}
\author[0000-0003-3843-3242]{Tibor~T\"or\"ok}
\affiliation{Predictive Science Inc., 9990 Mesa Rim Road, Suite 170, San Diego, CA 92121, USA}
\author[0000-0001-6590-3479]{Erika~Palmerio}
\affiliation{Predictive Science Inc., 9990 Mesa Rim Road, Suite 170, San Diego, CA 92121, USA}
\author[0000-0003-1759-4354]{Cooper~Downs}
\affiliation{Predictive Science Inc., 9990 Mesa Rim Road, Suite 170, San Diego, CA 92121, USA}
\author[0000-0001-7053-4081]{Viacheslav~S.~Titov}
\affiliation{Predictive Science Inc., 9990 Mesa Rim Road, Suite 170, San Diego, CA 92121, USA}
\author[0000-0002-4459-7510]{Mark~G.~Linton}
\affiliation{U.S.\ Naval Research Laboratory, 4555 Overlook Avenue, SW Washington, DC 20375, USA}
\author[0000-0002-2633-4290]{Ronald~M.~Caplan}
\affiliation{Predictive Science Inc., 9990 Mesa Rim Road, Suite 170, San Diego, CA 92121, USA}
\author[0000-0001-9231-045X]{Roberto~Lionello}
\affiliation{Predictive Science Inc., 9990 Mesa Rim Road, Suite 170, San Diego, CA 92121, USA}
\textit{}

\begin{abstract}
The trajectories of coronal mass ejections (CMEs) are often seen to substantially deviate from a purely radial propagation direction. Such deviations occur predominantly in the corona and have been attributed to ``channeling'' or deflection of the eruptive flux by asymmetric ambient magnetic fields. Here, we investigate an additional mechanism that does not require any asymmetry of the pre-eruptive ambient field. Using magnetohydrodynamic numerical simulations, we show that the trajectory of CMEs through the solar corona can significantly deviate from a radial direction when propagation takes place in a unipolar radial field. We demonstrate that the deviation is most prominent below ${\sim}15\,R_{\odot}$ and can be attributed to an ``effective ${\bf I} \times {\bf B}$ force'' that arises from the intrusion of a magnetic flux rope with a net axial electric current into a unipolar background field. These results are important for predictions of CME trajectories in the context of space weather forecasts, as well as for reaching a deeper understanding of the fundamental physics underlying CME interactions with the ambient fields in the extended solar corona.
\end{abstract}

\section{Introduction}
\label{s:intro}
%
Coronal mass ejections (CMEs) are huge expulsions of magnetized plasma from the Sun's atmosphere into interplanetary space, and the main driver of space-weather effects at Earth \cite[e.g.,][]{baker16,temmer21}. CMEs, and the often associated flares and prominence eruptions, are the largest transient energy-release events in the solar system and constitute different manifestations of a sudden and violent reconfiguration of the coronal magnetic field \citep[e.g.,][]{forbes00}. While there is a general agreement that CMEs and flares are powered by the free magnetic energy stored in current-carrying structures in the low corona, there exists an ongoing debate on the physical mechanisms that initiate and drive these events \citep[e.g.,][]{aulanier14,green18}. As CMEs expand and then propagate through the corona and the inner heliosphere, their interaction with the surrounding magnetic field and solar wind, or with other CMEs, can significantly change their properties \citep[e.g.,][]{lugaz17,manchester17}. The compression regions and shock waves created by sufficiently fast CMEs \citep[e.g.,][]{gopalswamy06} play an important role for generating gradual solar energetic particle (SEP) events \citep[e.g.,][]{kahler92,desai16}. Upon their arrival at Earth, CMEs can trigger geomagnetic storms as they interact with the terrestrial magnetosphere \citep[e.g.,][]{pulkkinen07,zhang.j21}. The resulting geoeffectiveness depends primarily on the CME's magnetic configuration, velocity, and ram pressure \citep[e.g.,][]{srivastava04,siscoe06}.

Whether or not a CME hits Earth depends first and foremost on the location of its source region on the Sun, but to a large extent also on its {\em trajectory}. While most CMEs propagate away from the Sun in a more or less radial direction, their trajectories can sometimes deviate from a radial propagation by several tens of degrees \citep[e.g.,][]{isavnin14}, which can lead to false space-weather alerts. A striking example is the January 7, 2014 event \citep[e.g.,][]{moestl15}. This fast CME (radial speed ${\sim}2500$\,km\,s$^{-1}$) originated in an active region (AR) close to the disk center and was accompanied by an X1.2-class flare. Based on its speed and origin, many forecasters predicted a transit time to Earth of ${\sim}36$ hours and a strong geomagnetic storm. However, the CME arrived significantly later (after about 49 hours), and it delivered only a glancing blow to Earth's magnetosphere, resulting in very minor geomagnetic activity \citep[see][]{mays15}. Therefore, in order to improve our capabilities of predicting space-weather effects associated with CMEs, one important element is to understand the circumstances and mechanisms that determine CME trajectories.

The deviation of CMEs from a radial trajectory has been often attributed to their deflection (a {\em change} of trajectory) at the typically stronger magnetic fields (and Alfv\'en velocities) of coronal holes (CHs) \citep[e.g.,][]{gopalswamy09,panasenco13}. Indeed, CMEs that are launched at high latitudes, close to a polar CH, are often seen to deflect toward the heliospheric current sheet (HCS) \citep[e.g.,][]{filippov01,kilpua09}, and this also has been found in numerical simulations \citep[e.g.,][]{zuccarello12,bemporad12,talpeanu22}. The amount of deflection seems to depend on several parameters, such as the CH area (width) and the CME speed \citep[e.g.,][]{xie09,mohamed12,wang.j.20a, wang.j.22a}, and such dependencies have recently been studied systematically in a series of numerical simulations \citep{sahade20,sahade21}. 

Simple analytical calculations of the restoring forces resulting from the compression of the ambient magnetic field by an expanding CME have shown that CMEs should be deflected toward regions of lower magnetic energy density \citep{shen.c11}, and this idea was successfully tested for a sample of eight CMEs by \citet{gui11}, although \cite{sierya20} found the opposite behavior in four of the 13 cases they studied. Using the forces acting on the surface of a CME draped by its ambient field, \citet{kay13, kay15a} developed a semi-analytical tool (ForeCAT) that models the deflection (and rotation) of CMEs during their propagation.

Non-radial CME trajectories have been also attributed to the presence of asymmetric magnetic fields in the CME source region, which lead to an immediate ``channeling'' or ``asymmetric expansion'' of the erupting flux \citep[e.g.,][]{liewer15,wang.r15,sahade22}. Indeed, a pre-eruptive magnetic flux rope (MFR) that is embedded in an asymmetric field adopts a tilted configuration \citep[see, e.g., Fig.\,4(c) in][]{titov14}. To first order, it can be expected that the erupting MFR will rise in the direction given by the tilt angle, as in the simulation shown in Fig.\,8 in \cite{torok13}. Such initial channeling was likely the main contributor to the strongly non-radial trajectory of the January 7, 2014 CME, since the source-region field was highly asymmetric \citep{moestl15}. Of course, CMEs that experience an initial non-radial rise may still be subjected to deflection as they travel away from their source region, sometimes resulting in a reversal of the propagation direction \citep[e.g.,][]{sahade23}.

These two mechanisms (channeling and deflection) appear to be the main contributors to the deviation of CME trajectories from a radial direction, as such deviations seem to take place predominantly in the corona, often below 5--10\,$R_\odot$ \citep{isavnin14, kay15b}. In interplanetary space, deviations from a radial propagation direction occur primarily due to the interaction with other CMEs \citep[e.g.,][]{xiong09,lugaz12}, but also via (the presumably relatively weak effect of) flux pile-up due to the presence of the Parker spiral \citep[e.g.,][]{gosling87b,wang.y04,zhuang19}.

In the investigation presented here we focus on an additional CME deflection mechanism, which, to the best of our knowledge, has been discussed so far only by \citet{lugaz11}. The mechanism is similar to the one proposed by \citet{isenberg07} to explain the rotation of erupting MFRs about their rise direction in the presence of an external shear field (oriented in the same direction as the initial MFR axis). The interaction of the axial MFR current, ${\bf I}_\mathrm{FR}$, with the shear field, ${\bf B}_\mathrm{EX}$, produces a Lorentz force, ${\bf F} = {\bf I}_\mathrm{FR} \times {\bf B}_\mathrm{EX}$, which points in opposite (horizontal) directions in the two MFR legs. The forces are balanced as long as the MFR remains in a stable magnetic equilibrium, but start to rotate the MFR's top part as soon as an eruption commences. This effect was modeled and studied systematically in \citet{kliem12} and \cite{zhou23}. 

In principle, the same mechanism can be applied to CMEs that propagate in a unipolar radial field \citep[see][]{lugaz11}. This will be the case, for instance, if the CME is launched below a pseudostreamer \citep[e.g.,][]{torok11} or outside the streamer belt \citep[e.g.,][]{liu.y07a}. Here the CME takes the role of the erupting MFR, while the radial field takes the role of the shear field. For a CME whose legs are connected to the solar surface, we then expect that its front part will be deflected out of the plane of propagation (since in the CME legs the axial MFR current will be approximately aligned with the radial field). We note that this may be analogous to the rotation of spheromaks in an ambient field, caused by the ``spheromak tilting instability'' (see \citealt{asvestari22} and references therein.)

An essential assumption made in \citet{isenberg07} is that the background shear field can freely penetrate the MFR body (see their Fig.~11(b)) and so produce the above Lorentz force, just as in the case of a  current-carrying conductor placed into a vacuum magnetic field. In a highly conducting solar plasma, however, the interaction between the MFR and the ambient field has a more intricate and mediated character. Intruding into a surrounding magnetized plasma, the erupting MFR preserves its internal field structure (as long as its axial field strength remains larger than the strength of the ambient field) by forming shielding current layers at its surface. For an MFR with a net axial (toroidal) current that intrudes into a unipolar radial field, this leads to a magnetic-pressure difference at the two flanks of the MFR (see \S\,\ref{s:results}) whose resultant force might, in principle, differ from the ${\bf I}_\mathrm{FR} \times {\bf B}_\mathrm{EX}$ force conjectured by \citet{isenberg07}. However, in the framework of an idealized 2D model, where the MFR is considered as a solid perfect conductor, \citet{yeh83} showed that this resultant force is formally equal to ${\bf I}_\mathrm{FR} \times {\bf B}_\mathrm{EX}$ if the original (unperturbed) ambient field is uniform, i.e., if the currents producing it are sufficiently far away from the MFR current, so that magnetic-pressure gradients in that field can be neglected \citep[otherwise, additional force terms come into play; see Eq.\,(15) in][]{yeh83}. It is important to note that the resultant force, which we refer to as the ``effective ${\bf I} \times {\bf B}$ force'' henceforth (to emphasize the formal resemblance to ${\bf I}_\mathrm{FR} \times {\bf B}_\mathrm{EX}$), does not require any asymmetry of the original background field. Consequently, a deflection of CMEs propagating in a radial field by this force should always be expected, {\em in addition} to the channeling/deflection caused by net restoring forces resulting from the presence of asymmetry in the original background field, as described above.

We note that, in addition to the approach by \cite{yeh83}, the surrounding magnetized plasma should be considered as a perfect conductor as well. This implies the formation of an additional current layer that tries to shield the background medium from the field produced by the MFR current, thereby decreasing the effective ${\bf I} \times {\bf B}$ force. However, as we shall see in \S\,\ref{s:results}, this effect is relatively weak, at least during the initial propagation of the CME, where most of its deflection takes place.

The goal of this paper is to investigate the deflection of CMEs by the effective ${\bf I} \times {\bf B}$ force in a systematic manner, using magnetohydrodynamic (MHD) simulations. We first consider, as a ``proof of concept'', a very idealized configuration in which the initial MFR field and background field are superimposed (\S\,\ref{s:poc}). We then employ a more realistic setup, in which CMEs are launched from bipolar ARs low in the corona before they encounter open-field regions, so that no initial superposition of the MFR and the with the large-scale, radial background field is present  (\S\,\ref{s:setup}). The results of this second investigation are presented in \S\,\ref{s:results}, which is followed by a discussion and our conclusions in \S\,\ref{s:discussion}. 

%
\begin{figure}[t]
\centering
\includegraphics[width=0.97\linewidth]{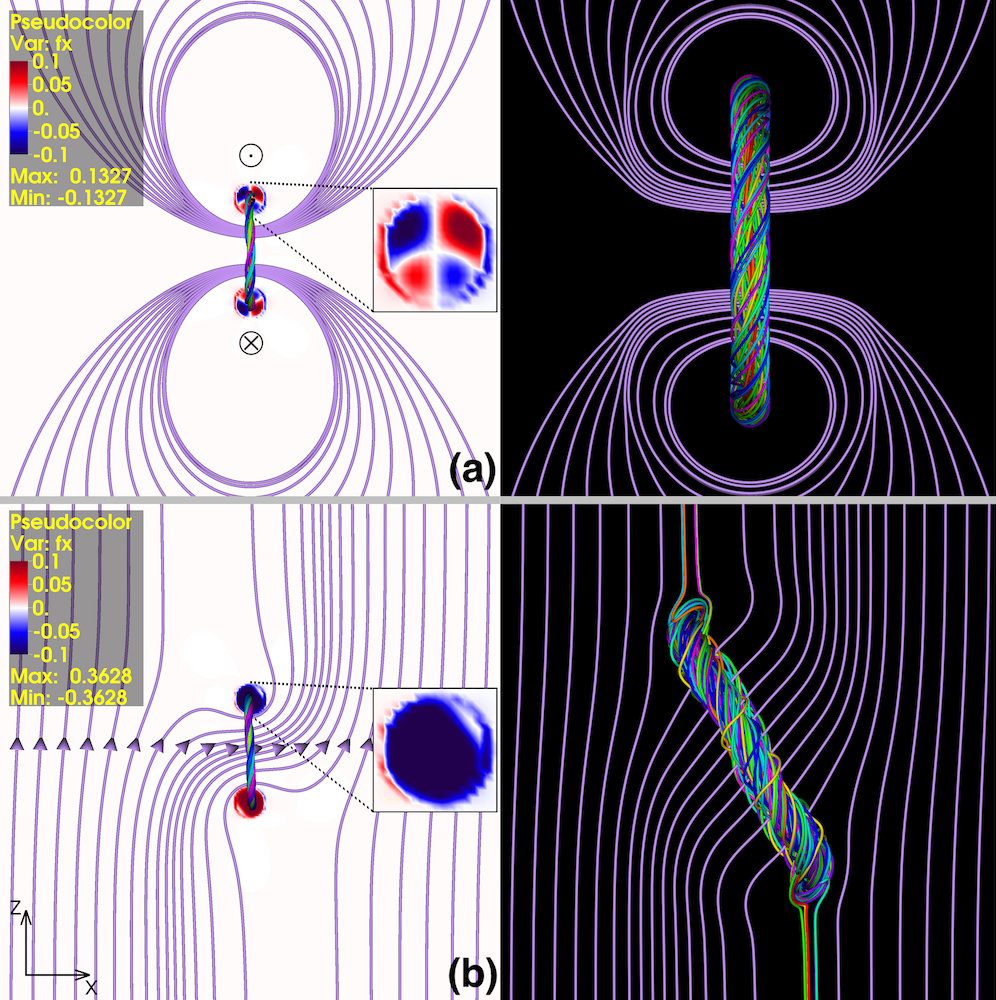}
\caption{
Simplified ``proof-of-concept'' numerical experiment. 
(a) Freely expanding toroidal MFR in a Cartesian box (side view, perpendicular to the torus axis). Shown are the initial state (left) and the MFR during its expansion (right). Rainbow-colored (purple) field lines show the MFR core (the field produced by the MFR current). The symbols denote the direction of the axial current. Red/blue colors show the horizontal component of the Lorentz force density, $\bf{j} \times \bf{B}$, in two MFR cross-sections. 
(b) Same initial MFR as in (a), now with a uniform vertical magnetic field, $B_z>0$, superimposed. This leads to a rotation of the MFR during its expansion. 
}
\label{f:fig1}
\end{figure}

\section{Simplified ``Proof-Of-Concept'' Experiment}
\label{s:poc}
%
Before we discuss our CME simulations in \S\ref{s:setup} and \S\ref{s:results}, we briefly describe here a simplified ``proof-of-concept'' experiment that we used to test the effect of a uniform ambient field on the evolution of an expanding MFR for a case where the MFR field and the ambient field are superimposed, as in \cite{isenberg07}. This experiment is shown in Figure\,\ref{f:fig1}. 

We consider the simplest possible configuration, namely a freely expanding current ring (a toroidal MFR) in a homogeneous background field. The simulation is performed in a Cartesian domain with the MHD code described in \citet{torok03}. We use the $\beta=0$ MHD approximation, where thermal pressure and gravity are neglected and the system is driven merely by inertial and Lorentz forces. The current ring is modeled by modifying the Titov--D{\'e}moulin MFR model \citep[TD;][]{titov99}, yielding thus the so-called ``modified Titov--D{\'e}moulin'' model \citep[TDm;][]{titov14}. In contrast to the TD model, which describes a partial current ring (a line-tied coronal MFR), we consider here a full current ring, and we remove the external fields that are present in the TD model. This means that the MFR is not in equilibrium and starts to expand right away at the beginning of the simulation. We place the MFR here such that the symmetry axis of the torus is parallel to the $x$ axis. 

Figure\,\ref{f:fig1}(a) shows a case without any external field. The left panel depicts the initial configuration. The inset shows that the horizontal Lorentz force densities along the symmetry axis of the torus are evenly distributed within the MFR. Consequently, as the MFR expands (right panel), it fully remains in its initial plane. Figure\,\ref{f:fig1}(b) shows a different experiment, where we add a uniform vertical field to the system. In the case shown, the strength of that field is 10\% of the maximum field strength within the MFR. We can see in the left panel that the initial horizontal Lorentz force densities are now oppositely directed in the top and bottom parts of the MFR, and that they point into the direction we expect from crossing the axial MFR current with the upward directed background field. Indeed, as the MFR expands, we now see a significant rotation (right panel), as expected from the considerations in \cite{isenberg07}.    

%
\begin{figure*}[t]
\centering
\includegraphics[width=0.822\linewidth]{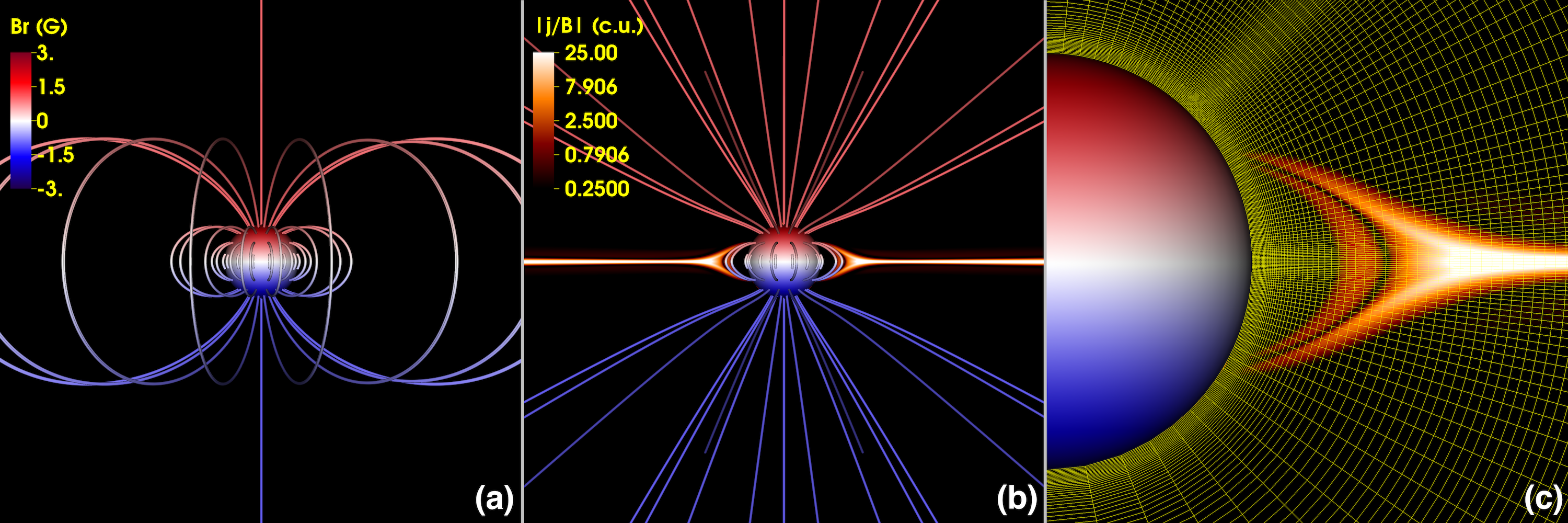}
\caption{
Global background magnetic configuration for our CME runs. 
(a) Initial configuration (global dipole). The sphere shows the solar surface, colored by the radial magnetic field, $B_r$. Only half of the simulation domain (out to about $15 R_\odot$) is shown here. Field lines are colored by $B_r/|{\bf B}|$. (b) Configuration after MHD relaxation, during which the solar wind opens up flux and the heliospheric current sheet (shown here in the plane $\phi=\pi$ by the quantity $|{\bf j}|/|{\bf B}|$) is formed. (c) Zoom into the streamer base, with the $(r,\theta)$ numerical mesh depicted in the plane $\phi=\pi$ (only every third grid point is shown in each direction). The mesh is finer at the Equator and at 45\degree  North, where an AR will be inserted.
}
\label{f:global_relaxed}
\end{figure*}

While this experiment demonstrates that the presence of a unipolar ambient field can indeed lead to a significant rotation of an expanding MFR, we have to keep in mind two strong simplifications. First, in our experiment the initial MFR and the vertical background field are superimposed , leading to strong sideways directed forces within the MFR. In a real CME, the pre-eruptive configuration (e.g., a line-tied MFR) is embedded in a closed, locally bipolar ambient field, which is often referred to as the ``strapping field''. Before the erupting MFR can encounter a unipolar radial field, it first has to overcome the strapping field, which may happen by pushing it to the side \citep[e.g.,][]{amari96}, via ``breakout'' reconnection above the MFR \citep[e.g.,][]{antiochos99,lynch08}, or via ``tether-cutting'' reconnection below the MFR that transfers strapping-field flux into flux of the MFR \citep[e.g.,][]{moore01}. Even then, as described in \S\,\ref{s:intro} and illustrated in Figure\,\ref{f:sketch_deflection} below, the background radial field cannot freely penetrate the MFR's magnetic field, due to the formation of shielding currents. Second, during the eruption the MFR will strongly expand, so its current and Lorentz force densities will be less concentrated than in our simplified experiment, where we inserted a compact MFR directly into a unipolar background field. Thus, in order to see whether a significant deflection still can take place after the initial MFR expansion and without a penetration of the MFR by the background field, we have to consider a model that includes the initial eruption that produces a CME. The model we are using for this task is described in the next section.    
 
%
\begin{figure*}[t]
\centering
\includegraphics[width=0.68\linewidth]{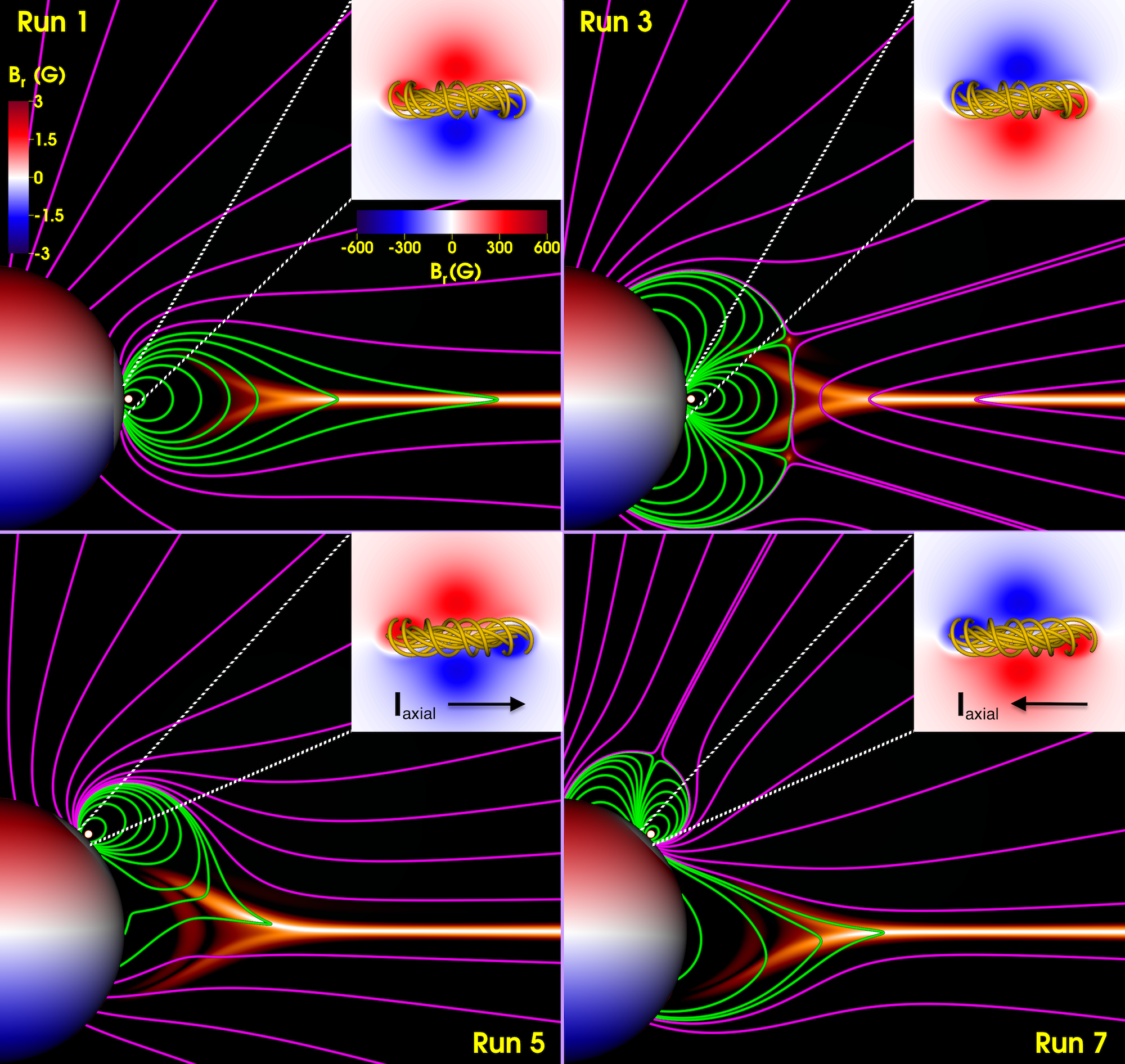}
\caption{
Initial configuration for Runs~1,~3,~5, and~7. The sphere shows $B_r$ at the solar surface; north/south is at the top/bottom. In Runs~1 and 3, the AR is placed below the streamer belt and HCS (cf. Figure\,\ref{f:global_relaxed}(b)), and in Runs~5 and~7 within an area of unipolar open field at 45\degree north. Green (closed) and magenta (open) field lines are shown, together with electric currents visualized by $|{\bf j}|/|{\bf B}|$) in the plane $\phi=\pi$ that cuts through the AR center. The insets show the AR magnetogram and the MFR (orange field lines) in a top-down view, where east/west is on the left/right. The MFR axial current (black arrows) is oppositely directed in the configurations on the left and right, respectively, because of the opposite orientations of the AR main polarities. All MFRs shown are right-handed (positive helicity). 
}
\label{f:fig_initial_config}
\end{figure*}

\section{Numerical Setup}
\label{s:setup}
%
Our simulations are similar to those described in \citet{lionello13}. They are performed with the Magnetohydrodynamic Algorithm Outside a Sphere (MAS) code, which solves the 3D time-dependent resistive MHD equations on a non-uniform spherical mesh \citep[see][and references therein]{torok18}.

Here we use MAS in the polytropic approximation \citep{linker99} with a fixed polytropic index of $\gamma=1.05$. For simplicity, and to better isolate the impact of the  effective ${\bf I} \times {\bf B}$ force, we do not consider additional waves that are able to accelerate a fast solar wind. This results in an overall slow solar wind profile that is roughly uniform in latitude. Our computational domain extends from $1\,R_\odot$ to $30\,R_\odot$ and is discretized on a non-uniform mesh in all three spherical coordinates ($r, \theta, \phi$), using $(405 \times 368 \times 318)$ points. The radial resolution of the mesh varies between $0.002\,R_\odot$ at the solar surface to $0.413\,R_\odot$ at the outer boundary.  

As described below, we use two different locations where we place an AR, namely either at the equator ($0^{\circ}$ latitude) or at $45^{\circ}$ north. To resolve the AR, we construct a finer mesh around these two latitudes. The minimum and maximum resolution of the mesh along longitudes (north--south; $\theta$) is $0.113\degree$ and $3.98\degree$, respectively, and along latitudes (east--west; $\phi$) it is $0.114\degree$ and $7.67\degree$, respectively. The resistivity and viscosity are chosen to be uniform and set such that the corresponding diffusion times are much larger than the Alfv\'en time of ${\approx}\,24$ minutes (by factors $10^{5}$ and $10^{3}$, respectively).

The background configuration of the corona starts from a global dipole field that is subjected to an MHD relaxation, during which the solar wind partially opens up the magnetic field. The resulting configuration (Figure~\ref{f:global_relaxed}) resembles solar minimum conditions, and at the solar surface the typical strength of the radial field is a few G. Following the relaxation we insert into this configuration a bipolar AR that contains an MFR in stable equilibrium, using the TDm model and two sub-photospheric magnetic charges to create the strapping field, as in the original TD model. Without loss of generality, we place the AR center at $\phi = \pi$. The maximum strength of the field in the AR is about 400~G. 

The insertion is done via superposition of the AR field and the background field. Since the AR field is two orders of magnitude larger than the background field, the strapping field spatially divides the MFR from the large-scale radial background field, into which the MFR intrudes only after its eruption. Note, also, that in the TDm model the MFR and the strapping field are superimposed in such a way that the initial net force acting on the MFR approximately vanishes (see \citealt{titov14} for details). Both aspects are different from the simple superposition of the MFR and background field in the configuration described in \S\,\ref{s:poc}. For simplicity, since we are only interested in the CME trajectory here, we do not modify the plasma parameters when inserting the AR.

We consider eight different configurations, which differ by the location of the inserted AR (equator or $45^{\circ}$ north, i.e., under the HCS or in a unipolar ambient field), its magnetic orientation (north--south or south--north), and the handedness of the MFR (right-handed or left-handed). In all eight configurations, the MFR axis is east--west oriented. The odd-numbered runs (Runs 1, 3, 5, and 7) and even-numbered runs (Runs 2, 4, 6, and 8) differ only in the direction of the MFR's axial current with respect to its axial field (the handedness), yielding a right-handed MFR for odd-numbered runs and a left-handed MFR for even-numbered runs (corresponding to positive and negative helicity, respectively). The initial configurations of the four runs with right-handed MFRs are shown in Figure\,\ref{f:fig_initial_config}. In Run~1 (top left panel), the AR is located below the HCS and it has the same magnetic orientation (north--south) as the global background field, resulting in a bipolar source region (see the closed green field lines). In Run~3 (top right panel), the AR orientation is reversed, resulting in a quadrupolar source region. In Run~5 (bottom left panel), the AR is inserted at $45^{\circ}$ north with the same magnetic orientation as in Run~1, resulting in a bipolar source region in which the closed field at larger heights is tilted towards the equator. Finally, in Run~7 (bottom right panel) the AR orientation is reversed again, resulting in a pseudostreamer-like source region, where the MFR is located in the southern lobe of the pseudostreamer.

We note that for these simulations we have specifically chosen to insert each combination of AR and MFR into the exact same background MHD solution to initiate the CME. Alternatively, we could have run eight separate MHD relaxations using the surface flux-distribution that results from the superposition of the AR, MFR, and global dipole for each case. Then, the current-carrying part of the MFR field could be embedded into the background configuration to initiate the CME without modifying $B_r$ at the surface \citep[e.g.,][]{torok18,titov22}. While the latter approach is desired for case-study modeling, doing so would yield a slight displacement of the HCS away from the equatorial plane in the relaxed background solution, which would formally be different for each case. The latitudinal displacement of the HCS would be small for Runs 1--4, but on the order of 8--14$^\circ$ for Runs 5--8 \citep[as estimated from PFSS extrapolations using the POT3D code;][]{caplan21}. This displacement would also be oppositely directed for Runs 5/6 (northward) and 7/8 (southward), respectively. Therefore, because our goal is to study the effective ${\bf I} \times {\bf B}$ force in a background that is both unipolar and as similar as possible from run to run, we opted for the simpler setup, where the respective AR-MFR configurations are inserted into the same dipolar global background.

\begin{figure}[t]
\centering
\includegraphics[width=1.\linewidth]{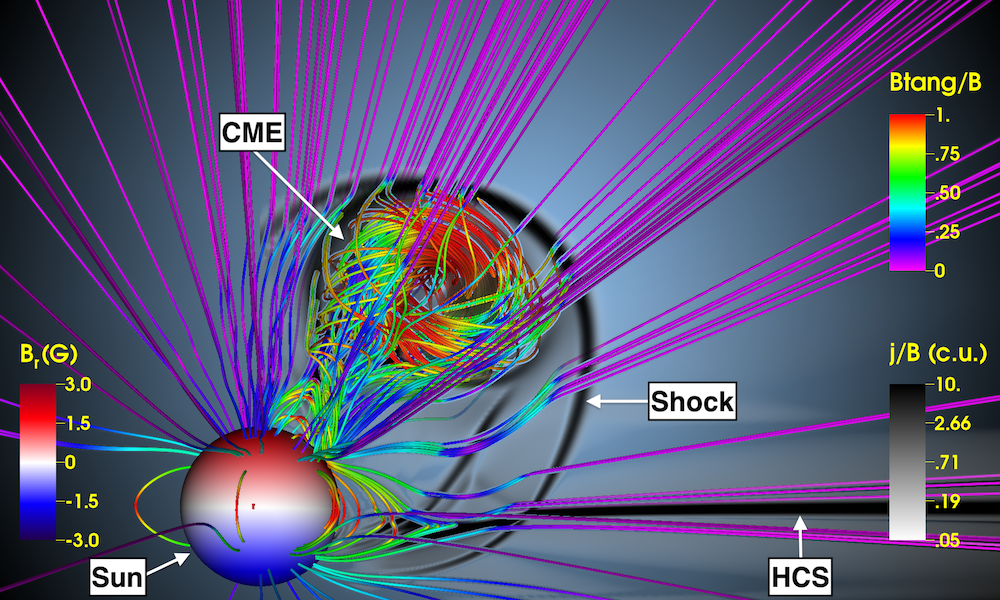}
\caption{
Erupting MFR after 3 Alfvén times (72 minutes) in Run~7. Field lines are colored by the normalized transverse magnetic field $(B_\theta^2 + B_\phi^2)^{1/2}/|\bf B|$ (denoted here as Btang/B). Electric currents are shown by $|{\bf j}|/|{\bf B}|$ in the $\phi=\pi$ plane, outlining the HCS and the shock wave produced by the CME.
}
\label{f:run07cme}
\end{figure}

For each of the eight configurations we perform a number of short test runs to determine the maximum MFR current that still yields a stable configuration. Once this value is determined, we trigger an eruption in each production run by setting the MFR current slightly (about $1\%$) above the stable-equilibrium value. This ensures that the eruption starts slowly and then accelerates, as observed for most CMEs \citep[e.g.,][]{vrsnak01}. Each simulation is run for 50 Alfv\'en times (about 20 hours). Figure\,~\ref{f:run07cme} shows, as an example, a snapshot of the erupting MFR (or CME) for Run~7.  

%

%
\begin{figure}[t]
\centering
\includegraphics[width=1.\linewidth]{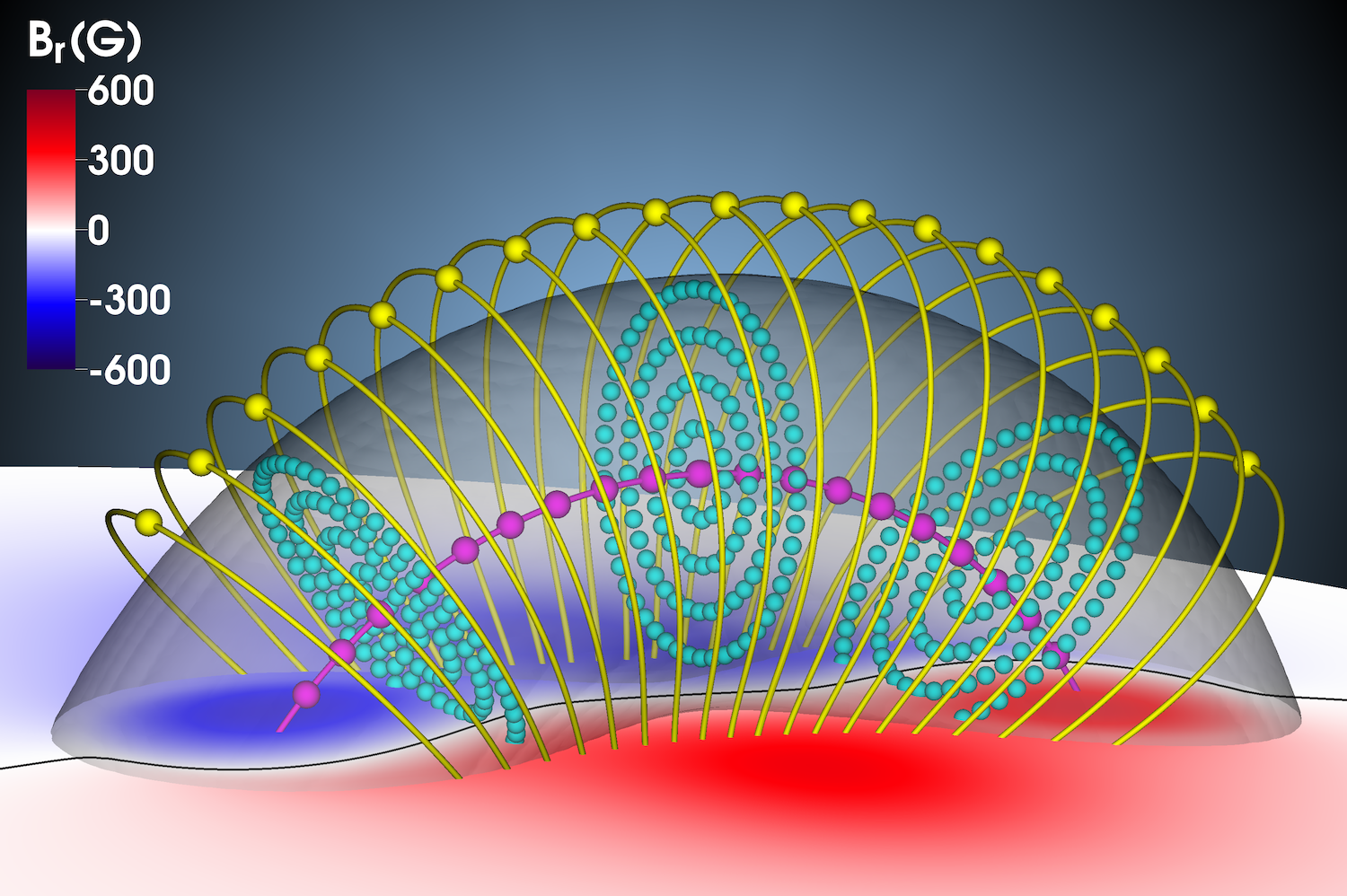}
\caption{
Distribution of tracer particles in and above the initial MFR (here for Run 7; cf. corresponding inset in Figure\,\ref{f:fig_initial_config}). For clarity, only selected tracers are displayed (as small spheres). Shown are 19 of the 1000 arc (yellow) and axis (magenta) tracers, and three of the 49 cross-sections (nos. 8, 25, and 42; in cyan) with 129 tracers each. Field lines originating at the tracers are shown for the arc and the axis. The transparent gray iso-surface shows $j=0.025\, j_\mathrm{max}$, outlining the edge of the MFR.     
}
\label{f:tracers}
\end{figure}

%
\begin{figure*}[t]
\centering
\includegraphics[width=0.67\linewidth]{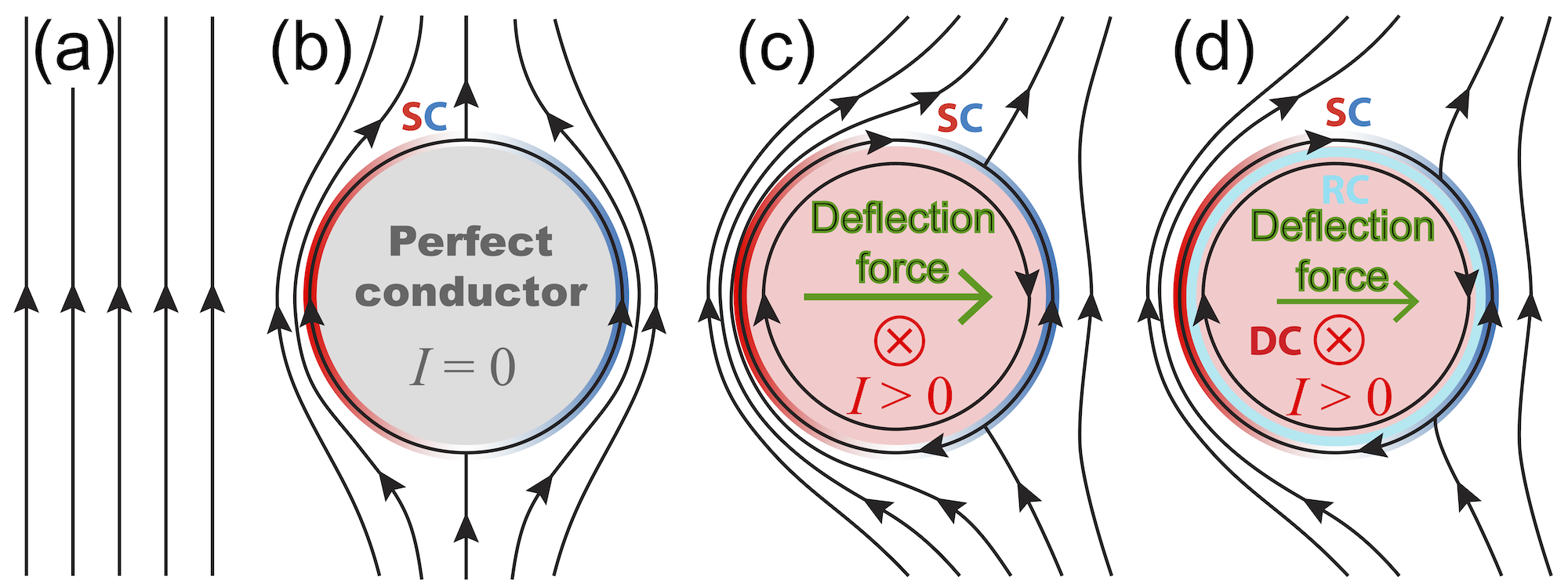}
\caption{
Sketch illustrating the effective ${\bf I} \times {\bf B}$ force introduced in \S\,\ref{s:intro}. (a) Unipolar background field. (b) The intrusion of a perfectly conducting MFR i.e., $I=0$, with a fully balancing ``direct'' and ``return'' axial current, not shown) leads to the formation of shielding surface currents (SC; shown in red and blue, where red depicts currents directed into the plane and blue depicts currents directed towards the viewer) and a deformation of the background field.  (c) An MFR with a direct current ($I>0$) additionally produces a magnetic-pressure difference in the ambient field, yielding a sideways deflection force. (d) Most general case with a return current (RC; light blue) and a direct current (DC), where the RC partially balances the DC. See text for further details.
}
\label{f:sketch_deflection}
\end{figure*}

At the beginning of each simulation, we distribute ``tracer particles'' (fluid elements) within and slightly above the MFR, and follow their trajectories over the course of the simulation. To do so, we interpolate, at each time-step and for each tracer, the velocities from the adjacent grid points to the tracer location, and advect the tracers according to the resulting velocity vector and current time step.  

Figure\,\ref{f:tracers} illustrates how the tracers are distributed initially. A first group of tracers is distributed evenly along the MFR axis. A second group consists of ``projections'' of each axis tracer along the radial direction, placed at a distance of 1.5 times the MFR minor radius away from the axis. This ``arc'' of tracers represents the closed arcade field located just above the MFR. Finally, a third group is distributed within evenly spaced cross-sections of the MFR (perpendicular to the axis). We place $1,000$ axis tracer particles, $1,000$ arc tracer particles and $129$ tracer particles at each of the MFR cross-sections (there are a total of $49$ such cross-sections). For our analysis described in the next section, we only include tracer particles that propagated in the radial direction at least 10\% above their initial position. We found that on average (for Runs~5--8, which are the runs that exhibit a deviation from a radial propagation; see below) $635$ axis and $436$ arc tracer particles fulfilled this condition, and all of these particles erupted as part of the CME. Therefore, we included five central cross-sections (within the vicinity of the MFR apex) in the analysis. This resulted in a similar number of $645$ cross-section tracer particles, which all erupted as well.

\section{Results}
\label{s:results}

\subsection{Effective ${\bf I} \times {\bf B}$ Force and Current Distribution}
\label{ss:currents}
%
Before discussing our simulation results, let us first consider in more detail the effective ${\bf I} \times {\bf B}$ force introduced in \S\,\ref{s:intro}. For our considerations, we assume the following properties of a CME propagating in a unipolar radial background field. First, the CME-MFR is sufficiently elongated for its magnetic structure to be approximately invariant along the axial direction. Second, over the course of its evolution, the MFR tends to preserve its integrity and approximately circular cross-sectional shape. Third, all currents remain concentrated within the MFR and its immediate neighborhood, so that the ambient magnetic field remains practically current-free. Fourth, adjacent magnetic field lines are largely tangential to the current-carrying volume. We note that these properties, which in large part are present in our simulations, allow one to determine the forces acting on the CME-MFR by integrating the Maxwell stress tensor over the surface of the current-carrying volume (rather than evaluating the Lorentz forces). We leave a quantitative analysis of the forces to a later study---here we merely employ the above properties to discuss the nature of the effective ${\bf I} \times {\bf B}$ force in a qualitative manner, as illustrated in Figure\,\ref{f:sketch_deflection}.

%
\begin{figure*}[t!]
\centering
\includegraphics[width=0.754\linewidth]{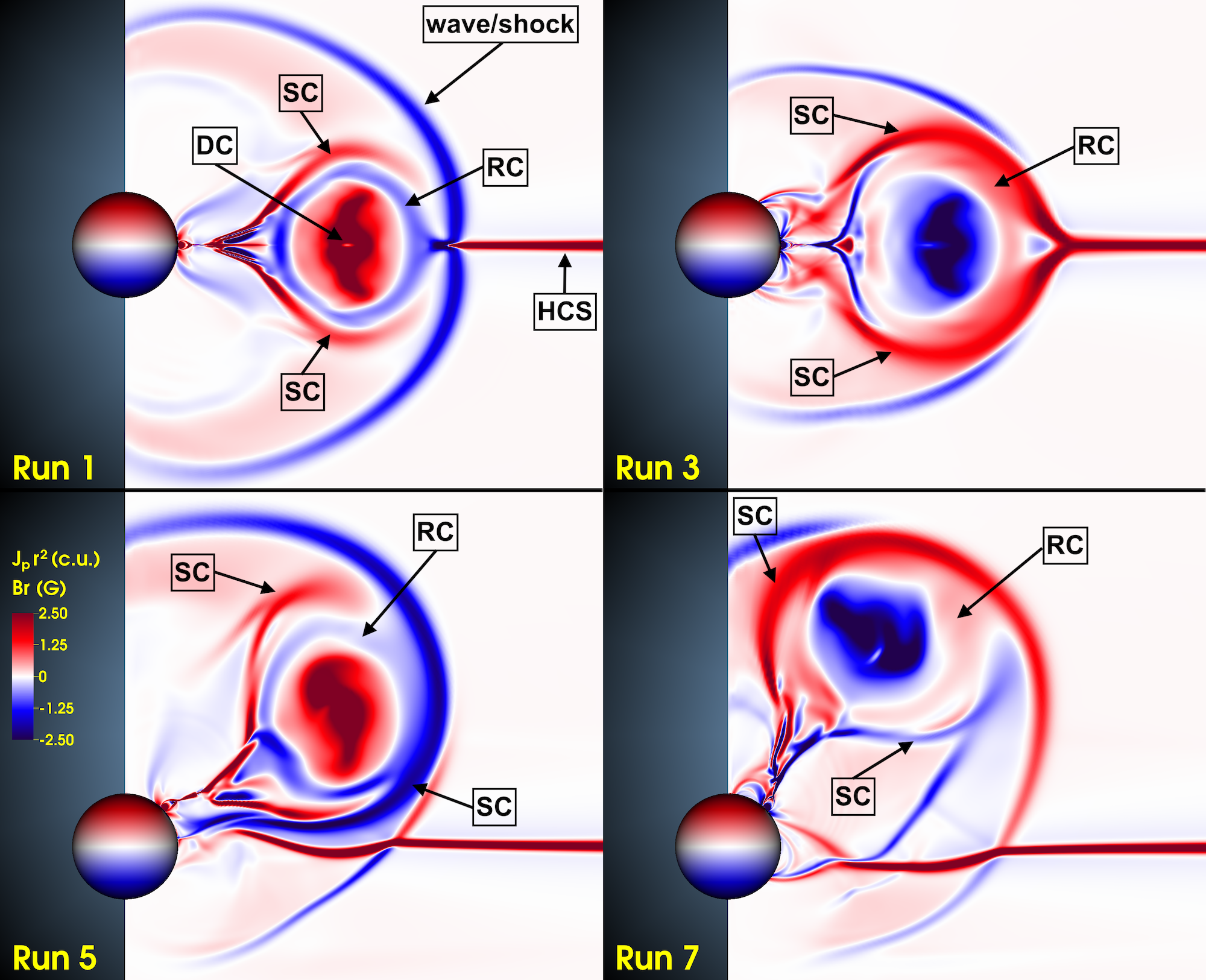}
\caption{
Distribution of electric currents in the $\phi=\pi$ plane for Runs 1, 3, 5, and 7 after 5 Alfvén times ($\approx$\,2 hours), visualized by $j_p\,r^2$ (in code units), where $j_p$ is the current density along the $\phi$ direction (which points into the plane) and $r$ is the (sun-centered) radial coordinate. Hence, red colors show currents directed into the plane, while blue colors show colors directed towards the viewer. ``HCS'', ``DC'', ``RC'', and ``SC'' stand for the heliospheric current sheet, direct (axial) current, return current, and shielding current, respectively.
}
\label{f:currents_t=205}
\end{figure*}

Consider, first, a unipolar background magnetic field as shown in panel (a). The intrusion of an MFR (in the simplest case, a perfect solid conductor, i.e., a fully current-neutralized MFR) into such a field implies the formation of oppositely directed surface shielding currents (SC) that prevent the background field from penetrating the MFR and force it to ``flow'' around it (panel (b)). However, if the MFR carries a direct current (DC), the poloidal (azimuthal) magnetic field produced by that current additionally contributes to the ambient field, increasing the tangential component of the ambient field on one side of the MFR and decreasing it on the other (panel (c)). This results in a sustained difference in magnetic pressure on the two flanks of the MFR, i.e., in a force that pushes the MFR sideways, with the oppositely directed field lines on the right-hand side facilitating magnetic reconnection (see \S\,\ref{s:discussion}). We refer to this sideward motion as ``deflection'' henceforth, but note that we do not mean here a change of direction due to a collision with some other structure, such as a CH. The direction of the deflection depends on the respective orientations of the initial background field and the MFR axial current. 

We note that the case shown in panel (c) has been considered earlier by \cite{yeh83}, who demonstrated that, for a uniform background field ${\bf B_0}$ and MFR axial current $I$, the corresponding net force exactly coincides with ${\bf I} \times {\bf B_0}$, as if $B_0$ would freely penetrate the MFR body as assumed by \cite{isenberg07} (see \S\,\ref{s:intro}). Our simulations described below suggest a more general case, which is sketched in panel (d) and includes a return current (RC) within the MFR. The presence of such a current can be understood if one considers that the magnetized plasma surrounding the MFR is an almost perfect fluid conductor, trying to prevent the azimuthal MFR field from penetrating the external volume by generating an RC. Note that the presence of an RC weakens the deflection force compared to the case shown in panel (c), but that force should still be present as long as the RC remains smaller than the direct current (DC) that flows in the opposite direction in the core of the MFR.   

Figure\,\ref{f:currents_t=205} shows the distributions of electric currents that develop during the eruption for Runs 1, 3, 5, and 7 (i.e., runs with a right-handed MFR). It can be seen that the DC, RC, and SC sketched in Figure\,\ref{f:sketch_deflection}(d) are present in each case. There are additional currents present in the system, predominantly at the wave/shock-front and in the wake of the CME, but those do not seem to be relevant for the CME deflection considered here. The sign of the DC in the MFR core (corresponding to $I_\mathrm{axial}$ in Figure\,\ref{f:fig_initial_config}) is determined by the magnetic orientation of the AR (see insets in Figure\,\ref{f:fig_initial_config}). It is, in all cases, surrounded by an RC of opposite sign near the periphery of the CME-MFR. The RC, in turn, is enclosed by the SC, whose sign is determined by the orientation of the ambient field. In Runs 1 and 3, where that field is oppositely directed on the two sides of the HCS, the SC has the same (positive) sign at both flanks of the CME. In Runs 5 and 7, on the other hand, the displaced radial field has the same sign at both flanks of the CME, so the SC has opposite signs at the respective CME flanks. The overall current distributions in the simulations are in line with the theoretical considerations outlined above. Hence, we expect to find a deflection of the CME by the effective ${\bf I} \times {\bf B}$ force in those runs where the CME propagates in a unipolar ambient field.

\begin{figure*}[t!]
\centering
\includegraphics[width=0.96\linewidth]{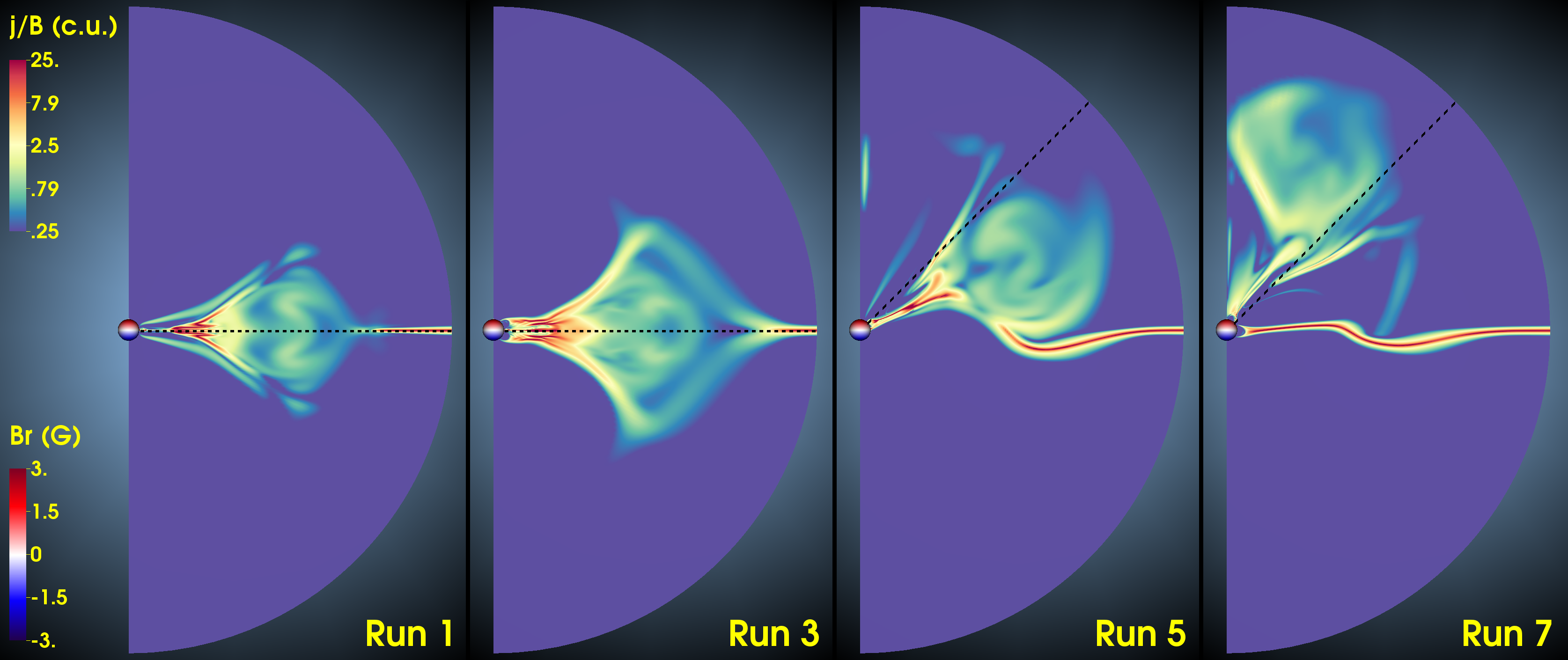}
\caption{
CMEs for Runs 1, 3, 5, and 7 after 25 Alfvén times ($\approx$10 hours), visualized by $|{\bf j}|/|{\bf B}|$ (in code units) in the $\phi=\pi$ plane. The dashed black line marks the radial direction above the AR. The radial extension of the simulation domain is $(1-30) R_\odot$. Runs 1 and 3 propagate along the HCS, while Runs~5 and~7 propagate in a unipolar radial ambient field (cf. Figure\,\ref{f:fig_initial_config}).
}
\label{f:j_over_b_t_225}
\end{figure*}

Figure\,~\ref{f:j_over_b_t_225} shows the CMEs for Runs 1, 3, 5, and 7, about ten hours after the CME is launched and before its front reaches the outer simulation boundary at $30\,R_\odot$. We can see that the CMEs in Runs~1 and 3, which are launched under the HCS, do not exhibit any signatures of deviation from a radial propagation. This is expected, since the effective $\bf{I}\times\bf{B}$ force points into opposite directions on the two sides of the HCS (due to the oppositely-directed radial background fields), so that the overall net effect is zero. 

In contrast, Runs 5 and 7, where the CME propagates into a purely unipolar (in this case, positive) radial field, exhibit a significant deviation from a radial propagation direction (indicated by the dashed black line). The respective directions of the deflection (southward for Run~5, northward for Run~7) are as expected from the effective $\bf{I}\times\bf{B}$ force for the respective orientations of the MFR axial currents (see Figure\,\ref{f:fig_initial_config}). 

%

As we describe below, our even-numbered runs behave essentially in the same way as the odd-numbered ones, except for the opposite sense of rotation of the erupting MFR about its rise direction. Therefore, we focus our analysis of CME deflection on Runs~5 and 7. Quantifying the deflection is not straightforward because the CME-MFRs do not expand in a simple, self-similar manner. We therefore consider the bulk {\em angular} CME deflection and employ two independent methods for estimating its amount.

\subsection{Tracer Particles}
\label{ss:deflection_tracers}
\begin{figure*}[t]
\centering
\includegraphics[width=1.0\linewidth]{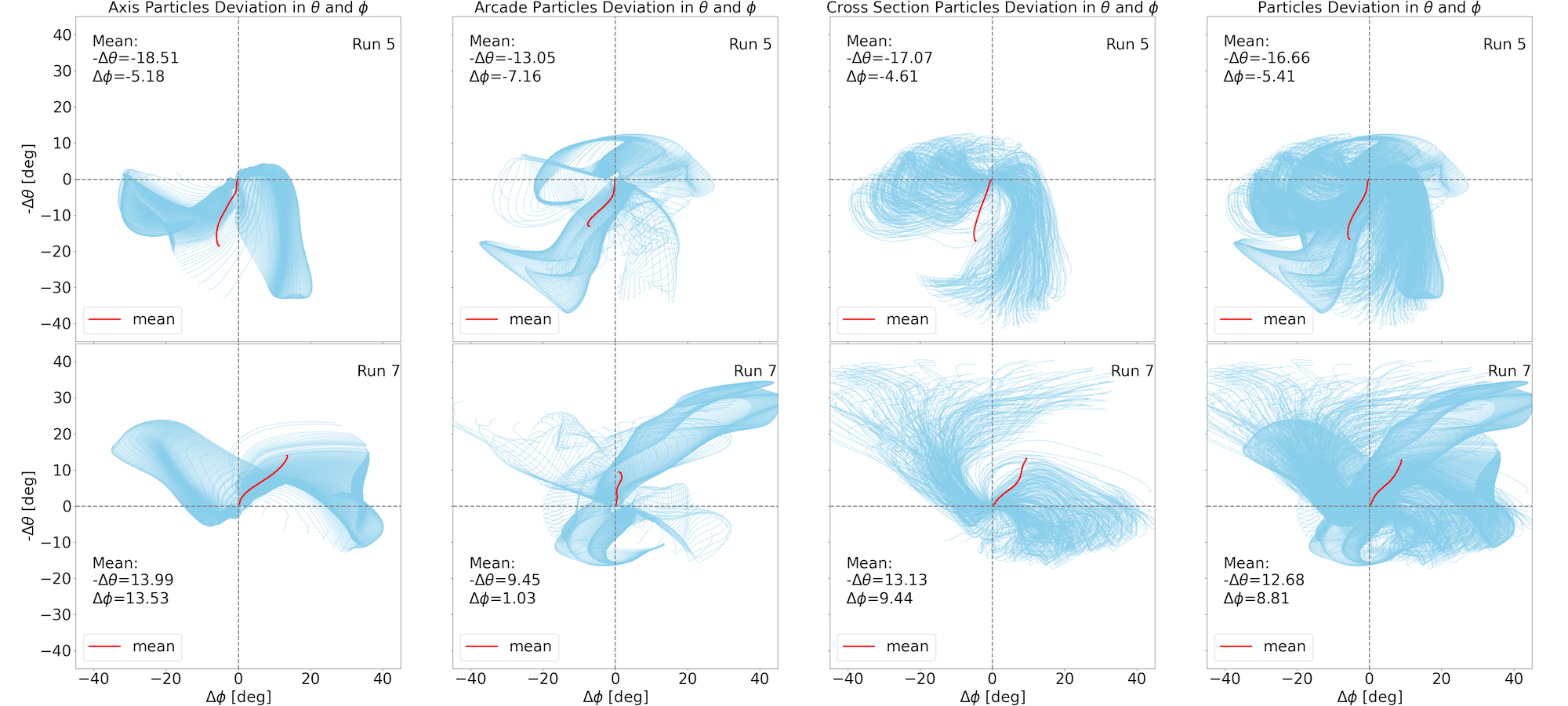}
\caption{
Trajectories of tracer particles in the $(\theta,\phi)$ plane (with respect to their initial positions in that plane) along the N--S ($-\Delta\theta$) and E--W ($\Delta\phi$) directions for Run~5 (top) and Run~7 (bottom). The mean trajectories, obtained from a center-of-mass calculation (see text for details) are shown in red. The first three panels show, respectively, tracers starting from the initial MFR axis, the arc above this axis, and five MFR cross-sections around the initial axis apex; the rightmost panel shows all tracers combined. The annotated values are the mean angular deviations at the time when the first tracer of the respective group reaches the outer boundary of the simulation domain. 
}
\label{f:axs_rng_dflct_run05_and_run07}
\end{figure*}

%
First, we employ our tracer particles to estimate the angular CME deflection. The first three panels in Figure\,\ref{f:axs_rng_dflct_run05_and_run07} depict, for comparison, the angular deviations in spherical space for the three groups of tracers described in \S\,\ref{s:setup} (for each group, we follow the tracers' positions until the first tracer hits the outer boundary at 30\,$R_\odot$). Shown are the east--west (E--W; along constant latitudes) and north--south (N--S; along constant longitudes) deviations of individual tracers (light-blue curves) over time for Run~5 and Run~7 (top and bottom rows, respectively). The rightmost panel shows the full set of tracers included in our analysis. 

We associate the bulk angular deflection of the CME with the angular deflection of the center of mass of the full set of tracers. To obtain the latter, we compute their mean position in Cartesian coordinates, and convert that position back to spherical coordinates. The red curves in the rightmost panel of Figure\,\ref{f:axs_rng_dflct_run05_and_run07} show the resulting evolution of the center of mass over the course of the simulation. For completeness, we also show the corresponding values for our three tracer groups separately.

For Run~5, we thus find a total (final) southward and eastward deflection of 17\degree and 5\degree, respectively, while for Run~7 we find a total northward and westward deflection of 13\degree and 9\degree, respectively. The bottom panel in Figure\,\ref{f:gcs_evolution} below shows the evolution of the CME deflection in the N--S direction over time for both runs. In Runs~1 and 3, where there is no deflection, the corresponding angular tracer deviations (not shown here) merely represent the lateral expansion of the CME, with all mean values being zero.

%

We note that the selected tracers represent the evolving CME volume fairly well, but not completely. Specifically, they lag behind the visible CME leading edge, which has already left the simulation domain when the first tracers arrive at $r=30\,R_\odot$. This discrepancy is because of the so-called stand-off distance between the actual CME-MFR (as covered by the tracers) and the ``observed'' CME leading edge, i.e., the shock/wave front (see \citealt{ma11}, \citealt{gopalswamy12}, and Figure\,\ref{f:gcs_snapshots} below). We also note that the arc tracers, which are initially located entirely above the MFR, largely become part of the CME-MFR, presumably due to reconnection occurring between the expanding MFR and adjacent flux systems \citep[e.g.,][]{aulanier19}. This justifies the inclusion of those tracers for our deflection analysis.
 
Due to the asymmetric distribution of the closed-field regions above the AR in Runs~5 and 7 (see green field lines in Figure\,\ref{f:fig_initial_config}), some deflection along the N-S direction already occurs {\em before} the MFR reaches the purely radial field at a height of less than 2\,$R_\odot$. However, as can be seen in Figure\,\ref{f:gcs_evolution} (bottom), this initial phase amounts only to a relatively small fraction of the total deflection (it takes place way before the first respective data point in the figure). At even lower heights, on the scale of the AR, the field is almost perfectly symmetric, since the AR field is two orders of magnitude larger than the ambient field. Hence, during the very early phase of the MFR eruption the rise direction should be almost perfectly radial.

While the deflections in N--S (or S--N) direction are expected from the respective direction of the effective $\bf{I}\times\bf{B}$ force, the cause of the (mostly significantly smaller) deflections in E--W (or W--E) direction is less obvious. They most likely result from the bulk rotation of the top part of the erupting MFR about its rise direction (see Figure\,\ref{f:run07cme}), which locally changes the orientation of the axial MFR current and, hence, the direction of the effective $\bf{I}\times\bf{B}$ force. Indeed, the eastward (westward) deflection in Run~5 (Run~7) is in line with what one would expect from the sense of rotation \citep[clockwise for right-handed MFRs; e.g.,][]{green07,lynch09,kliem12} and the resulting change of direction of the effective $\bf{I}\times\bf{B}$ force. Moreover, for our Runs~6 and 8 (which differ from Runs~5 and 7 only in the MFR handedness) we find the same southward and northward, but exactly opposite westward and eastward deflection, respectively, which concurs with the opposite sense of rotation (counter-clockwise) in those runs. Such rotations about the rise direction are expected for erupting MFRs even in the absence of an external shear field, in which case they are caused by the conversion of MFR twist into writhe \citep[e.g.,][]{torok10}. 

%
\begin{figure*}[t!]
\centering
\includegraphics[width=0.91\linewidth]{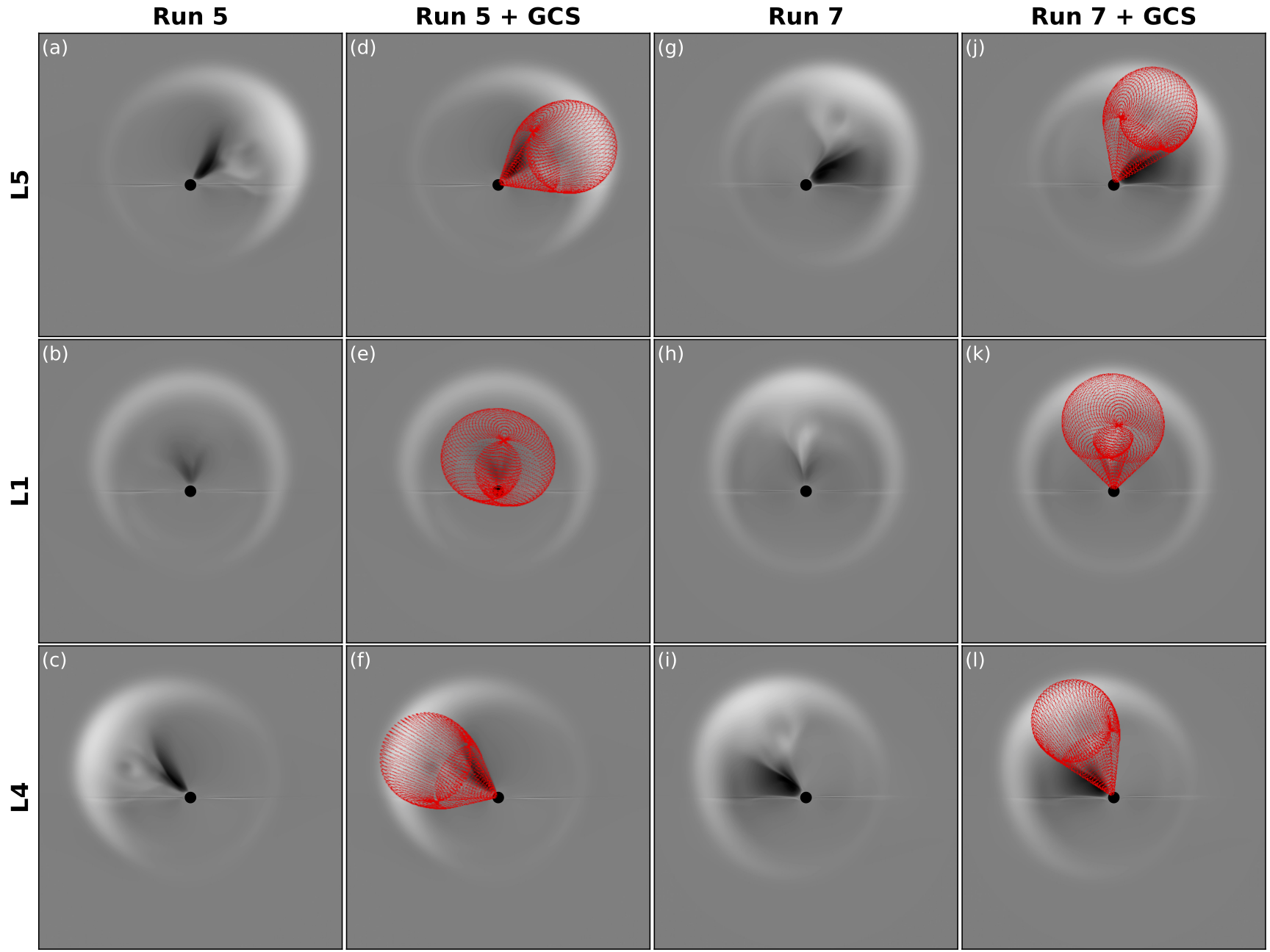}
\caption{
Example of GCS fitting results for (a--f) Run~5 and (g--l) Run~7. Panels (a--c) and (g--h) show base-difference synthetic coronagraph images (where the frame at $t=0.0$~hours is subtracted from the frame at $t=8.8$~hours) from the simulated L5, L1, and L4 points, while panels (d--f) and (j--l) show the same images with the GCS wireframe projection (in red) overlaid.
}
\label{f:gcs_snapshots}
\end{figure*}

\subsection{GCS Modeling}
\label{ss:deflection_GCS}
%
The second method we use for determining the CME deflection is based on a technique that is widely employed to characterize CME propagation through the solar corona for real events, namely forward-modeling via the Graduated Cylindrical Shell \citep[GCS;][]{thernisien11} model. In brief, the GCS model consists of a parametrized (hollow) shell, with a half-toroidal main body and two conical legs connected back to the Sun (reminiscent of an MFR morphology). Forward-modeling with the GCS technique is performed by visually fitting the shell to simultaneous coronagraph images, while tuning its six free parameters ($\theta$, $\phi$, nose height, axial tilt, center-to-leg half-width, and aspect ratio) until the wireframe's projection onto each plane of sky best matches the observational data. This technique has been employed to evaluate CME deflection through the corona in a number of studies \citep[e.g.,][]{gui11,kay17}. In order to apply the GCS model to our simulated CMEs, synthetic coronagraph images need to be produced. Theoretically, there are no constraints on where to place synthetic observers for the generation of such images but, in order to mimic a likely real-event situation, we choose to set three viewing perspectives at a heliocentric distance of 1~au and along the solar equatorial plane. One observer is placed at $\phi=180^{\circ}$, i.e.\ aligned in longitude with the source AR, and the remaining two are placed at $\phi=120^{\circ}$ and $\phi=240^{\circ}$, simulating the Sun--Earth L1, L5, and L4 points, respectively. SOHO and the twin STEREO spacecraft (all three equipped with coronagraphs) were in this approximate configuration in October 2009; furthermore, the L4 and L5 points have gained considerable attention in recent years because of their potential benefits for space-weather forecasting \citep[e.g.,][]{vourlidas15,bemporad21,posner21}. Once the synthetic observers have been set, we produce corresponding coronagraph images (i.e., total K-corona brightness maps; see \citealt{mikic94} and \citealt{howard09} for a detailed description of the procedure) throughout the simulation's temporal (${\sim}20$~hours) and spatial ($[1$--$30]\,R_{\odot}$) domains. 

We perform GCS fittings of the CME for both Run~5 and Run~7, at regular intervals (in time) until the CME's nose/apex exits the coronagraph's field of view in at least one data set, for a total of 10 fits. An example of such fits, performed at $t=8.8$~hours, is shown in Figure\,\ref{f:gcs_snapshots}. It is evident from these snapshots that the CMEs in Run~5 and Run~7 have already deflected considerably southward and northward, respectively, at this point in the simulation (cf. Figure\,\ref{f:j_over_b_t_225}). By $t=9.8$~hours and at a GCS estimated apex height of ${\sim}29\,R_{\odot}$, just before the CME leading edge reaches the outer boundary, we find a total deflection in latitude of $17^{\circ}$ southward for Run~5 and $15^{\circ}$ northward for Run~7 with the GCS method.

As for the E--W direction, our GCS fitting results do not exhibit significant deflections, with values within ${\pm}2^{\circ}$ of the AR longitude ($\phi=180^{\circ}$). This is different from the tracer-particles method (see \S\,\ref{ss:deflection_tracers}), and may be due to the larger uncertainties that are known to be associated with estimates of $\phi$ in comparison to $\theta$---\citet{thernisien09} performed a sensitivity analysis of the GCS model parameters, finding a mean deviation of ${\pm}1.8^{\circ}$ in $\theta$ and ${\pm}4.3^{\circ}$ in $\phi$---indicating that a deflection of just a few degrees in $\phi$ is unlikely to be recovered reliably via the GCS technique.

\subsection{Method Comparison \& Deflection Evolution}
\label{ss:deflection_evolution}
\begin{figure}[t!]
\centering
\includegraphics[width=1.\linewidth]{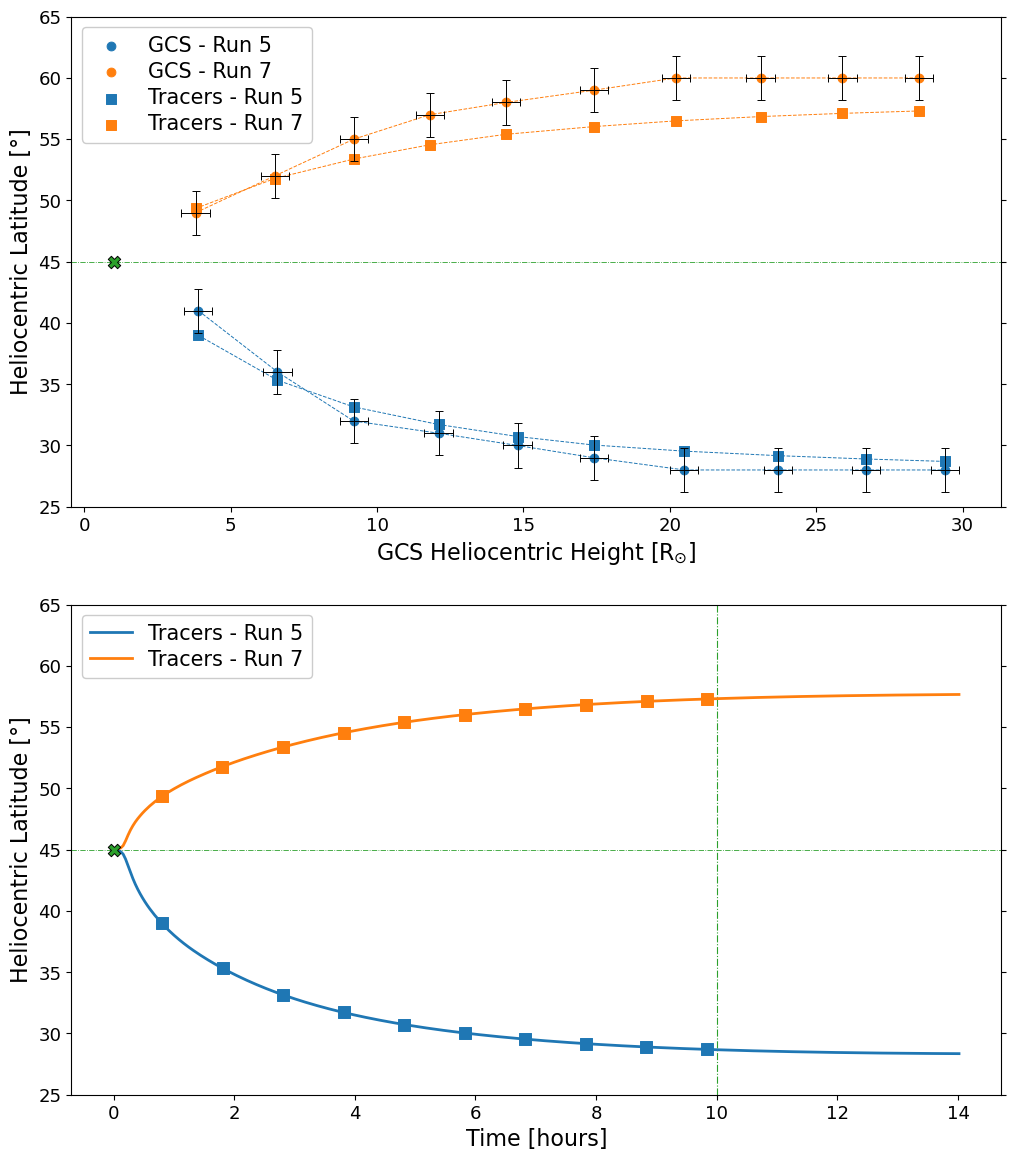}
\caption{
{\em Top:} Heliocentric latitude at the position of the CME leading-edge height (as determined from the GCS model apex) for Run~5 (blue) and Run~7 (orange). The circles and squares denote the latitudes estimated from the GCS fitting method and the tracer-particles method, respectively, at corresponding times. The green cross marks the location of the source AR at $45^{\circ}$ north. For the GCS fitting method, the error bars indicate the estimates for the average deviations in latitude (${\pm}1.8^{\circ}$) and height (${\pm}0.48\,R_{\odot}$), found by \citet{thernisien09}. 
{\em Bottom:} Heliocentric latitude but as a function of time from the tracer-particles method. The blue and orange lines use all the tracer-particle data (up to 14 hours), while the squares correspond to the data points shown in the top panel. The green vertical line marks the arrival time of the CME leading edge at the outer boundary of the simulation domain. 
}
\label{f:gcs_evolution}
\end{figure}

%
The evolution of the CME deflection across the whole simulation domain is shown as a function of height for both runs and methods in the top panel of Figure\,\ref{f:gcs_evolution}. For comparison, we plot in the bottom panel the complete temporal evolution as obtained from the tracer-particles method. Here we consider only the dominant deflection along the N--S direction. The plots show that, for both runs, the strongest deflection occurs within a few solar radii above the surface, which is in line with previous observational studies \citep[e.g.,][]{isavnin14,kay15b}. Subsequently, the added deflection successively decreases, and there are no more significant changes beyond $\approx 20\,R_\odot$. For Run~5 (blue) the GCS and tracer-particles methods yield consistent deflection angles, except for the first data point, where the value obtained from the tracer-particle method is slightly outside of the estimated GCS error bars. For Run~7 the correspondence is not quite as good: starting with the fourth data point, the GCS method consistently yields a somewhat larger deflection angle, with a difference of 2\degree\ for the final data point. 

%

%
\begin{figure}[t!]
\centering
\includegraphics[width=1.\linewidth]{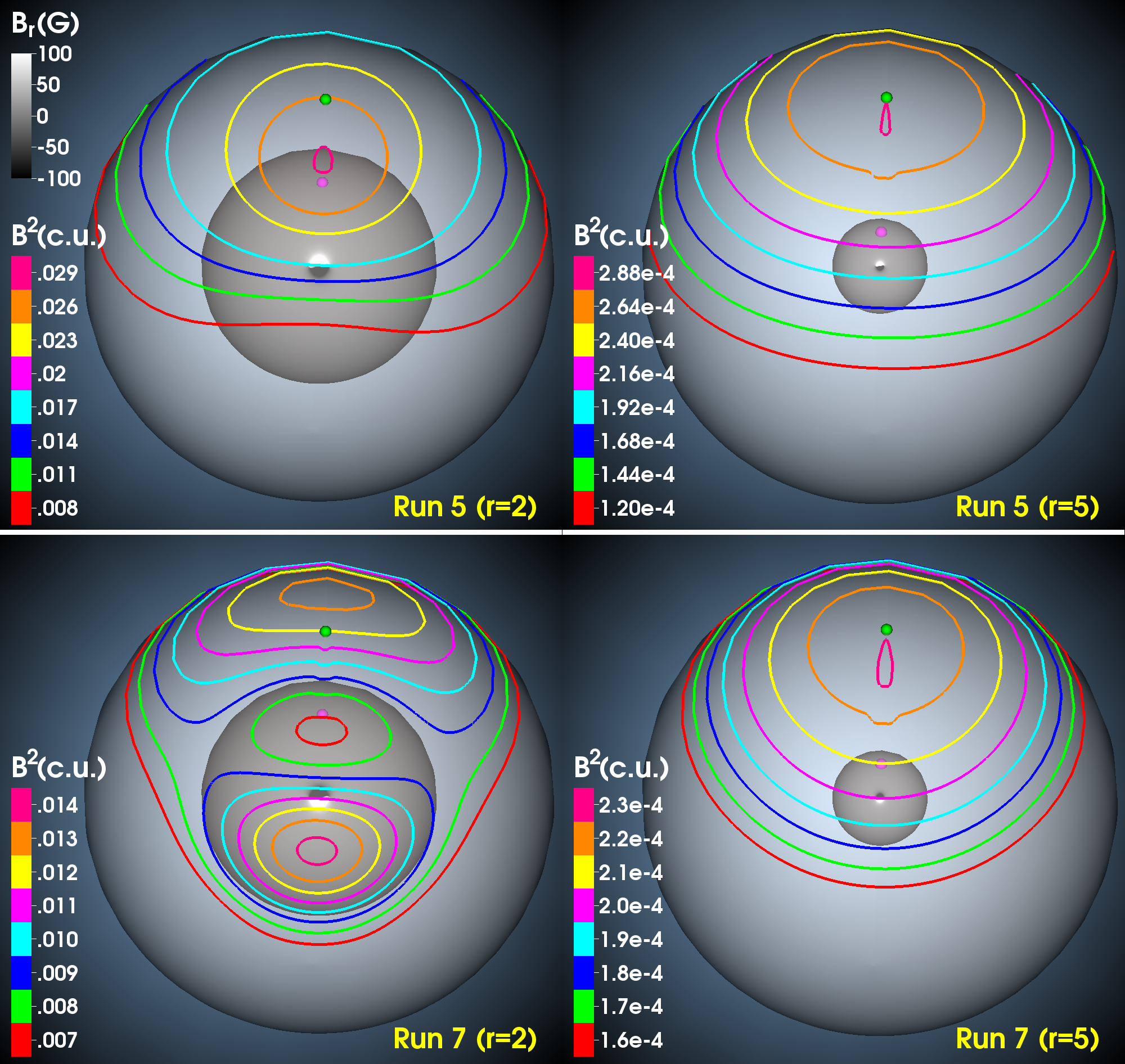}
\caption{
Contours of $B^2$ (in code units) for Runs~5 (top) and~7 (bottom) at 2 (left) and 5 (right) $R_\odot$ for the total magnetic field (including the AR) at the beginning of the simulation. The inner sphere shows the radial magnetic field (in gray-scale) at the solar surface. The view is centered on the AR at 45\degree north. The small magenta (green) ball marks the position of the solar north pole (its radial projection onto the respective outer sphere).
}
\label{f:b2_contour}
\end{figure}

By inspecting the morphology of the CMEs in both runs in synthetic white-light data (Figure\,\ref{f:gcs_snapshots}), we see that the CME in Run~5 is characterized by a relatively symmetric overall structure, while the CME in Run~7 displays stronger asymmetries, with its main body ``bending'' northward with respect to the underlying signature of the current sheet (see also Figure\,\ref{f:j_over_b_t_225}). The GCS shell, however, does not allow for such deformations, making the fitting procedure more difficult in the case of CMEs with irregular morphologies. Additionally, since the GCS forward-modeling method is performed ``manually''---or ``by eye''---it follows that there are intrinsic uncertainties related to the specific user's fitting choices. The effects of the user's subjectivity on GCS results have been explored and discussed in \citet{verbeke22}. Taking these considerations into account, we deem the results obtained with the tracer-particles method to be more robust.

Looking again at Figure\,\ref{f:gcs_evolution}, we see that the respective amounts of deflection differ significantly between Run~5 and Run~7. To understand this, we have to consider the influence of the global ambient field on the CME trajectory. Figure\,\ref{f:b2_contour} shows the initial (i.e., after the AR has been inserted) distribution of the (normalized) magnetic energy density, $B^2$, of the total magnetic field at two representative coronal heights. For Run~5 (top panels), $B^2$ decreases monotonically from the north pole to the equator at both heights, i.e., the net restoring force acting due to the compression of the ambient field by the CME \citep{shen.c11} always points southward, supporting the southward directed effective $\bf{I}\times\bf{B}$ force at all times. For Run~7, however, the effective $\bf{I}\times\bf{B}$ force is northward directed, so the net restoring force acts {\em against} it, at least at coronal heights where the global field dominates (bottom right panel). For lower heights (bottom left panel), the situation is more complicated than in Run~5, due to the opposite magnetic orientation of the AR. Here $B^2$ has a local minimum at a latitude higher than that of the AR, so initially the restoring force should support the northward deflection by the effective $\bf{I}\times\bf{B}$ force. Overall, however, the counteracting effect of the net restoring force dominates, leading to a smaller total deflection than in Run~5.

\section{Summary and Discussion}
\label{s:discussion}
%
We used MHD simulations to systematically investigate the deflection of CMEs in a unipolar coronal magnetic field, focusing here on the deflection caused by the ``effective $\mathbf{I} \times \mathbf{B}$ force'' described in \S\S\,\ref{s:intro} and \ref{s:results}. Our ``proof of concept'' simulation described in \S\,\ref{s:poc} showed that a freely expanding MFR remains in its initial plane, but the addition (superposition) of a uniform vertical magnetic field leads to a rotation of the MFR during its expansion, as suggested by \cite{isenberg07}. However, this simulation ignores one important condition for real CMEs, namely that the ambient field cannot freely penetrate the CME-MFR. Moreover, the axial MFR current remains much more compact than in a real CME, where the MFR undergoes a strong expansion.

We addressed these limitations by constructing a global coronal configuration that resembles solar-minimum conditions, consisting of a purely unipolar field in each hemisphere and an HCS in the equatorial plane. Into this background configuration we inserted a bipolar AR that contained an MFR. We created eight configurations that differed in the location of the AR (under the HCS or the radial unipolar field at 45\degree,N), its magnetic orientation (N--S or S--N), and the MFR handedness (right- or left-handed). For each configuration we initiated an eruption of the MFR and quantified the CME deflection using two independent methods, namely a set of tracer particles (fluid elements) and GCS modeling. 

We found that CMEs launched under the HCS did not exhibit any deviation from a radial propagation, due to the cancellation of deflection forces across the HCS. On the other hand, CMEs launched into the (in our case positive) unipolar radial field exhibited a significant southward or northward angular deflection, depending on the AR orientation, i.e., the direction of the axial MFR current. While the (relatively modest) asymmetry of the closed field overlying the initial MFR likely leads to some channeling (i.e., additional deflection) of the MFR in the early phase of the eruption, our results indicate that the bulk of the CME's deflection is accrued during its propagation in the open radial field.

Both the GCS technique and the tracer-particles method estimate the total southward deflection (Run~5) to be $17^{\circ}$. For the total northward deflection (Run~7) the GCS technique estimate was $15^{\circ}$, while the tracer-particles method gave a somewhat lower estimate of $\mathbf{13^{\circ}}$. We attributed the latter difference to limitations of the GCS method for asymmetric CME geometries. As discussed in \S\,\ref{ss:deflection_evolution}, the fact that the northward deflection is smaller than the southward one is consistent with the northward-pointing magnetic pressure gradient of the large-scale coronal background field. The resulting, southward-directed net restoring force (due to the compression of that field by the CME) increases the deflection of the southward moving CME, whereas it counteracts the deflection of the northward moving one. The substantial total northward deflection in Run~7 suggests that the effective ${\bf I} \times {\bf B}$ force is significantly stronger than the net restoring force for the configurations we considered here. Overall, our results are in line with those found by \citet{lugaz11}, who reported a CME deflection by the effective ${\bf I} \times {\bf B}$ force of $10^{\circ}$ (at $5.0\,R_\odot$) and $11.5^{\circ}$ (at $8.3\,R_\odot$) in an MHD simulation of an observed event. 

For the CMEs that propagate in the unipolar radial ambient field we also found a (smaller) deflection in the E--W (or W--E) direction, which we attributed to changes of the orientation of the axial current due to the bulk rotation of the erupting MFR about its rise direction. For MFRs with the same initial orientation of the axial current but opposite handedness (i.e., opposite sense of rotation), we found the exactly same amount, but opposite direction of that deflection. This indicates that the initial MFR twist, a main factor in determining the amount of rotation, influences the partition of the deflection into longitudinal and latitudinal components. The CMEs launched under the HCS, while undergoing a rotation as well, do not exhibit any eastward or westward deflection due to force cancellation across the HCS.

We note that the expansion and propagation of the CMEs in our simulations is accompanied by continuous reconnection between the CME-MFR and the ambient field. This occurs, for example, between the axial field of the MFR legs and the ambient field in the form of ``interchange reconnection'' \citep[e.g.,][]{reinard04, lugaz11, crooker12}. For our deflected CMEs, which propagate in a unipolar radial ambient field, interchange reconnection is dominant in the CME-MFR leg in which the MFR axial field is largely antiparallel to the ambient field. Over time, this leads to the ``opening'' of a considerable fraction of the MFR flux. This partially changes the path of the axial MFR current and should, therefore, have some effect on the CME deflection by the effective ${\bf I} \times {\bf B}$ force. We expect reconnection also to take place between the azimuthal MFR flux and the ambient field via ``breakthrough reconnection'' \citep{titov22}, which is analogous to ``breakout reconnection'' \citep{antiochos99}, except that it can occur in bipolar magnetic configurations. As described in \cite{titov22}, breakthrough reconnection takes place at quasi-separatrix layers \citep[QSLs;][]{priest95,demoulin96}. In our cases of CME propagation in a unipolar ambient field, this QSL is expected to form on the flank of the CME-MFR, where the MFR azimuthal field is directed opposite to the ambient field, i.e., on the right-hand side of the MFR in Figure\,\ref{f:sketch_deflection}(c,d). Reconnection on this side of the MFR then ``peels away'' adjacent ambient field, allowing the compressed flux on the other side of the MFR to push the CME more efficiently. Quantifying the exact locations of reconnection, the respective amounts of magnetic fluxes transferred, and the role of breakthrough reconnection for the CME deflection, are nontrivial tasks, which we leave for future work.

As shown in Figure\,\ref{f:gcs_evolution}, the deflection continuously decelerates as the CME moves away from the Sun, until there is no more notable increase beyond $\approx 20\,R_\odot$. The reason for this saturation of the CME deflection is not obvious. It may take place, for instance, because other forces such as the aerodynamic drag force or the magnetic tension force in the MFR legs start to counterbalance the effective ${\bf I} \times {\bf B}$ force and/or because the strength of the MFR return current with respect to the direct current increases over time until it fully neutralizes the latter, so that the effective ${\bf I} \times {\bf B}$ force vanishes. Such effects are not straightforward to quantify, as they require a detailed assessment of the current-evolution and the various forces involved, so we leave also this question for a future investigation.

Follow-up work will also explore several model parameters that have not been varied here. For example, we expect the ratio of the strength of the AR/MFR field to the strength of the large-scale radial field (assumed to be ${\sim}100$ in the pre-eruptive configuration in this study) to have an effect on the total deflection. Additionally, the orientation of the source-region polarity inversion line, i.e, of the initial MFR (in our case strictly E--W or W--E) may play a role, as well as the amount by which the erupting MFR rotates about its rise direction, which can be controlled to some extent by, e.g., the initial MFR diameter \citep[][]{kliem12}. Finally, future work will examine more closely the early evolution (0--5 hours) of the erupting MFR, as this interval corresponds to the range of coronal heights (${<}15\,R_{\odot}$) where most of the deflection was observed to take place.

The simulations presented in this work demonstrate the significance of the effective ${\bf I} \times {\bf B}$ force for the development of non-radial CME trajectories, in addition to initial flux channeling or deflection by asymmetric source-region or ambient magnetic fields. The effective ${\bf I} \times {\bf B}$ force should act most efficiently on CMEs propagating in a largely unipolar radial ambient field. That is, it seems most relevant for CMEs launched below a pseudostreamer or outside the streamer belt, and presumably during periods of lower solar activity, when the HCS is less complex and larger coherent areas of unipolar field should exist at low and middle latitudes. Its relevance  could be tested, for instance, with future Parker Solar Probe \citep{fox16} in-situ observations of CMEs below ${\sim}20\,R_{\odot}$. 

Furthermore, our results could be applied to predictions of CME trajectories throughout the extended solar corona, with direct implications for the so-called ``CME hit/miss problem'' in space weather forecasting \citep[e.g.,][]{wold18, verbeke19}. This could be achieved, for example, by incorporating the effective $\mathbf{I} \times \mathbf{B}$ force (or, more generally, the concept of shielding currents that develop in response to CME-MFRs intruding non-uniform ambient fields, as outlined in \citealt{yeh83}) into (semi-)empirical models used for such predictions.

\section*{Acknowledgments}
We thank the anonymous referee for their patience and very detailed and helpful comments that significantly improved the quality of this article. The work described here was supported by NASA's HSR, HSO-Connect, and HTMS programs (award No.\ 80NSSC20K1317, 80NSSC20K1285, and 80NSSC20K1274, respectively) and by NSF's PREEVENTS and Solar--Terrestrial programs (award No.\ ICER-1854790 and AGS-1923377, respectively). T.T.\ was additionally supported by Naval Research Laboratory (NRL) contract N0017319C2003 to Predictive Science Inc., a subcontract of NASA/LWS grant 80HQTR19T0084 to NRL. M.G.L.\ acknowledges support by the latter contract and by the Office of Naval Research.

\bibliographystyle{yahapj}
\bibliography{bennun_al}

\begin{thebibliography}{}
\providecommand\natexlab[1]{#1}
\providecommand\JournalTitle[1]{#1}

\bibitem[{{Amari} {et~al.}(1996){Amari}, {Luciani}, {Aly}, \&
  {Tagger}}]{amari96}
{Amari}, T., {Luciani}, J.~F., {Aly}, J.~J., \& {Tagger}, M. 1996,
  \href{http://dx.doi.org/10.1086/310158}{\JournalTitle{\apjl}, 466, L39}

\bibitem[{{Antiochos} {et~al.}(1999){Antiochos}, {DeVore}, \&
  {Klimchuk}}]{antiochos99}
{Antiochos}, S.~K., {DeVore}, C.~R., \& {Klimchuk}, J.~A. 1999,
  \href{http://dx.doi.org/10.1086/306563}{\JournalTitle{\apj}, 510, 485}

\bibitem[{{Asvestari} {et~al.}(2022){Asvestari}, {Rindlisbacher}, {Pomoell}, \&
  {Kilpua}}]{asvestari22}
{Asvestari}, E., {Rindlisbacher}, T., {Pomoell}, J., \& {Kilpua}, E. K.~J.
  2022, \href{http://dx.doi.org/10.3847/1538-4357/ac3a73}{\JournalTitle{\apj},
  926, 87}

\bibitem[{{Aulanier}(2014)}]{aulanier14}
{Aulanier}, G. 2014, \href{http://dx.doi.org/10.1017/S1743921313010958}{in
  Nature of Prominences and their Role in Space Weather, ed. B.~{Schmieder},
  J.-M. {Malherbe}, \& S.~T. {Wu}, Vol. 300}, 184

\bibitem[{{Aulanier} \& {Dud{\'\i}k}(2019)}]{aulanier19}
{Aulanier}, G., \& {Dud{\'\i}k}, J. 2019,
  \href{http://dx.doi.org/10.1051/0004-6361/201834221}{\JournalTitle{\aap},
  621, A72}

\bibitem[{{Baker} \& {Lanzerotti}(2016)}]{baker16}
{Baker}, D.~N., \& {Lanzerotti}, L.~J. 2016,
  \href{http://dx.doi.org/10.1119/1.4938403}{\JournalTitle{\amjph}, 84, 166}

\bibitem[{{Bemporad}(2021)}]{bemporad21}
{Bemporad}, A. 2021,
  \href{http://dx.doi.org/10.3389/fspas.2021.627576}{\JournalTitle{\frass}, 8,
  627576}

\bibitem[{{Bemporad} {et~al.}(2012){Bemporad}, {Zuccarello}, {Jacobs},
  {Mierla}, \& {Poedts}}]{bemporad12}
{Bemporad}, A., {Zuccarello}, F.~P., {Jacobs}, C., {Mierla}, M., \& {Poedts},
  S. 2012,
  \href{http://dx.doi.org/10.1007/s11207-012-9999-3}{\JournalTitle{\solphys},
  281, 223}

\bibitem[{{Caplan} {et~al.}(2021){Caplan}, {Downs}, {Linker}, \&
  {Mikic}}]{caplan21}
{Caplan}, R.~M., {Downs}, C., {Linker}, J.~A., \& {Mikic}, Z. 2021,
  \href{http://dx.doi.org/10.3847/1538-4357/abfd2f}{\JournalTitle{\apj}, 915,
  44}

\bibitem[{{Crooker} \& {Owens}(2012)}]{crooker12}
{Crooker}, N.~U., \& {Owens}, M.~J. 2012,
  \href{http://dx.doi.org/10.1007/s11214-011-9748-1}{\JournalTitle{\ssr}, 172,
  201}

\bibitem[{{Demoulin} {et~al.}(1996){Demoulin}, {Henoux}, {Priest}, \&
  {Mandrini}}]{demoulin96}
{Demoulin}, P., {Henoux}, J.~C., {Priest}, E.~R., \& {Mandrini}, C.~H. 1996,
  \JournalTitle{\aap}, 308, 643

\bibitem[{{Desai} \& {Giacalone}(2016)}]{desai16}
{Desai}, M., \& {Giacalone}, J. 2016,
  \href{http://dx.doi.org/10.1007/s41116-016-0002-5}{\JournalTitle{\lrsp}, 13,
  3}

\bibitem[{{Filippov} {et~al.}(2001){Filippov}, {Gopalswamy}, \&
  {Lozhechkin}}]{filippov01}
{Filippov}, B.~P., {Gopalswamy}, N., \& {Lozhechkin}, A.~V. 2001,
  \href{http://dx.doi.org/10.1023/A:1012754329767}{\JournalTitle{\solphys},
  203, 119}

\bibitem[{{Forbes}(2000)}]{forbes00}
{Forbes}, T.~G. 2000,
  \href{http://dx.doi.org/10.1029/2000JA000005}{\JournalTitle{\jgr}, 105,
  23153}

\bibitem[{{Fox} {et~al.}(2016){Fox}, {Velli}, {Bale}, {Decker}, {Driesman},
  {Howard}, {Kasper}, {Kinnison}, {Kusterer}, {Lario}, {Lockwood}, {McComas},
  {Raouafi}, \& {Szabo}}]{fox16}
{Fox}, N.~J., {Velli}, M.~C., {Bale}, S.~D., {et~al.} 2016,
  \href{http://dx.doi.org/10.1007/s11214-015-0211-6}{\JournalTitle{\ssr}, 204,
  7}

\bibitem[{{Gopalswamy}(2006)}]{gopalswamy06}
{Gopalswamy}, N. 2006,
  \href{http://dx.doi.org/10.1007/BF02702527}{\JournalTitle{\japa}, 27, 243}

\bibitem[{{Gopalswamy} {et~al.}(2009){Gopalswamy}, {M{\"a}kel{\"a}}, {Xie},
  {Akiyama}, \& {Yashiro}}]{gopalswamy09}
{Gopalswamy}, N., {M{\"a}kel{\"a}}, P., {Xie}, H., {Akiyama}, S., \& {Yashiro},
  S. 2009, \href{http://dx.doi.org/10.1029/2008JA013686}{\JournalTitle{\jgr},
  114, A00A22}

\bibitem[{{Gopalswamy} {et~al.}(2012){Gopalswamy}, {Nitta}, {Akiyama},
  {M{\"a}kel{\"a}}, \& {Yashiro}}]{gopalswamy12}
{Gopalswamy}, N., {Nitta}, N., {Akiyama}, S., {M{\"a}kel{\"a}}, P., \&
  {Yashiro}, S. 2012,
  \href{http://dx.doi.org/10.1088/0004-637X/744/1/72}{\JournalTitle{\apj}, 744,
  72}

\bibitem[{{Gosling} {et~al.}(1987){Gosling}, {Thomsen}, {Bame}, \&
  {Zwickl}}]{gosling87b}
{Gosling}, J.~T., {Thomsen}, M.~F., {Bame}, S.~J., \& {Zwickl}, R.~D. 1987,
  \href{http://dx.doi.org/10.1029/JA092iA11p12399}{\JournalTitle{\jgr}, 92,
  12399}

\bibitem[{{Green} {et~al.}(2007){Green}, {Kliem}, {T{\"o}r{\"o}k}, {van
  Driel-Gesztelyi}, \& {Attrill}}]{green07}
{Green}, L.~M., {Kliem}, B., {T{\"o}r{\"o}k}, T., {van Driel-Gesztelyi}, L., \&
  {Attrill}, G.~D.~R. 2007,
  \href{http://dx.doi.org/10.1007/s11207-007-9061-z}{\JournalTitle{\solphys},
  246, 365}

\bibitem[{{Green} {et~al.}(2018){Green}, {T{\"o}r{\"o}k}, {Vr{\v{s}}nak},
  {Manchester}, \& {Veronig}}]{green18}
{Green}, L.~M., {T{\"o}r{\"o}k}, T., {Vr{\v{s}}nak}, B., {Manchester}, W., \&
  {Veronig}, A. 2018,
  \href{http://dx.doi.org/10.1007/s11214-017-0462-5}{\JournalTitle{\ssr}, 214,
  46}

\bibitem[{{Gui} {et~al.}(2011){Gui}, {Shen}, {Wang}, {Ye}, {Liu}, {Wang}, \&
  {Zhao}}]{gui11}
{Gui}, B., {Shen}, C., {Wang}, Y., {et~al.} 2011,
  \href{http://dx.doi.org/10.1007/s11207-011-9791-9}{\JournalTitle{\solphys},
  271, 111}

\bibitem[{{Howard} \& {Tappin}(2009)}]{howard09}
{Howard}, T.~A., \& {Tappin}, S.~J. 2009,
  \href{http://dx.doi.org/10.1007/s11214-009-9542-5}{\JournalTitle{\ssr}, 147,
  31}

\bibitem[{{Isavnin} {et~al.}(2014){Isavnin}, {Vourlidas}, \&
  {Kilpua}}]{isavnin14}
{Isavnin}, A., {Vourlidas}, A., \& {Kilpua}, E.~K.~J. 2014,
  \href{http://dx.doi.org/10.1007/s11207-013-0468-4}{\JournalTitle{\solphys},
  289, 2141}

\bibitem[{{Isenberg} \& {Forbes}(2007)}]{isenberg07}
{Isenberg}, P.~A., \& {Forbes}, T.~G. 2007,
  \href{http://dx.doi.org/10.1086/522025}{\JournalTitle{\apj}, 670, 1453}

\bibitem[{{Kahler}(1992)}]{kahler92}
{Kahler}, S.~W. 1992,
  \href{http://dx.doi.org/10.1146/annurev.aa.30.090192.000553}{\JournalTitle{\araa},
  30, 113}

\bibitem[{{Kay} {et~al.}(2017){Kay}, {Gopalswamy}, {Xie}, \& {Yashiro}}]{kay17}
{Kay}, C., {Gopalswamy}, N., {Xie}, H., \& {Yashiro}, S. 2017,
  \href{http://dx.doi.org/10.1007/s11207-017-1098-z}{\JournalTitle{\solphys},
  292, 78}

\bibitem[{{Kay} \& {Opher}(2015)}]{kay15b}
{Kay}, C., \& {Opher}, M. 2015,
  \href{http://dx.doi.org/10.1088/2041-8205/811/2/L36}{\JournalTitle{\apjl},
  811, L36}

\bibitem[{{Kay} {et~al.}(2013){Kay}, {Opher}, \& {Evans}}]{kay13}
{Kay}, C., {Opher}, M., \& {Evans}, R.~M. 2013,
  \href{http://dx.doi.org/10.1088/0004-637X/775/1/5}{\JournalTitle{\apj}, 775,
  5}

\bibitem[{{Kay} {et~al.}(2015){Kay}, {Opher}, \& {Evans}}]{kay15a}
{Kay}, C., {Opher}, M., \& {Evans}, R.~M. 2015,
  \href{http://dx.doi.org/10.1088/0004-637X/805/2/168}{\JournalTitle{\apj},
  805, 168}

\bibitem[{{Kilpua} {et~al.}(2009){Kilpua}, {Pomoell}, {Vourlidas}, {Vainio},
  {Luhmann}, {Li}, {Schroeder}, {Galvin}, \& {Simunac}}]{kilpua09}
{Kilpua}, E.~K.~J., {Pomoell}, J., {Vourlidas}, A., {et~al.} 2009,
  \href{http://dx.doi.org/10.5194/angeo-27-4491-2009}{\JournalTitle{\angeo},
  27, 4491}

\bibitem[{{Kliem} {et~al.}(2012){Kliem}, {T{\"o}r{\"o}k}, \&
  {Thompson}}]{kliem12}
{Kliem}, B., {T{\"o}r{\"o}k}, T., \& {Thompson}, W.~T. 2012,
  \href{http://dx.doi.org/10.1007/s11207-012-9990-z}{\JournalTitle{\solphys},
  281, 137}

\bibitem[{{Liewer} {et~al.}(2015){Liewer}, {Panasenco}, {Vourlidas}, \&
  {Colaninno}}]{liewer15}
{Liewer}, P., {Panasenco}, O., {Vourlidas}, A., \& {Colaninno}, R. 2015,
  \href{http://dx.doi.org/10.1007/s11207-015-0794-9}{\JournalTitle{\solphys},
  290, 3343}

\bibitem[{{Linker} {et~al.}(1999){Linker}, {Miki{\'c}}, {Biesecker}, {Forsyth},
  {Gibson}, {Lazarus}, {Lecinski}, {Riley}, {Szabo}, \& {Thompson}}]{linker99}
{Linker}, J.~A., {Miki{\'c}}, Z., {Biesecker}, D.~A., {et~al.} 1999,
  \href{http://dx.doi.org/10.1029/1998JA900159}{\JournalTitle{\jgr}, 104, 9809}

\bibitem[{{Lionello} {et~al.}(2013){Lionello}, {Downs}, {Linker},
  {T{\"o}r{\"o}k}, {Riley}, \& {Miki{\'c}}}]{lionello13}
{Lionello}, R., {Downs}, C., {Linker}, J.~A., {et~al.} 2013,
  \href{http://dx.doi.org/10.1088/0004-637X/777/1/76}{\JournalTitle{\apj}, 777,
  76}

\bibitem[{{Liu}(2007)}]{liu.y07a}
{Liu}, Y. 2007, \href{http://dx.doi.org/10.1086/511385}{\JournalTitle{\apjl},
  654, L171}

\bibitem[{{Lugaz} {et~al.}(2011){Lugaz}, {Downs}, {Shibata}, {Roussev}, {Asai},
  \& {Gombosi}}]{lugaz11}
{Lugaz}, N., {Downs}, C., {Shibata}, K., {et~al.} 2011,
  \href{http://dx.doi.org/10.1088/0004-637X/738/2/127}{\JournalTitle{\apj},
  738, 127}

\bibitem[{{Lugaz} {et~al.}(2012){Lugaz}, {Farrugia}, {Davies}, {M{\"o}stl},
  {Davis}, {Roussev}, \& {Temmer}}]{lugaz12}
{Lugaz}, N., {Farrugia}, C.~J., {Davies}, J.~A., {et~al.} 2012,
  \href{http://dx.doi.org/10.1088/0004-637X/759/1/68}{\JournalTitle{\apj}, 759,
  68}

\bibitem[{{Lugaz} {et~al.}(2017){Lugaz}, {Temmer}, {Wang}, \&
  {Farrugia}}]{lugaz17}
{Lugaz}, N., {Temmer}, M., {Wang}, Y., \& {Farrugia}, C.~J. 2017,
  \href{http://dx.doi.org/10.1007/s11207-017-1091-6}{\JournalTitle{\solphys},
  292, 64}

\bibitem[{{Lynch} {et~al.}(2008){Lynch}, {Antiochos}, {DeVore}, {Luhmann}, \&
  {Zurbuchen}}]{lynch08}
{Lynch}, B.~J., {Antiochos}, S.~K., {DeVore}, C.~R., {Luhmann}, J.~G., \&
  {Zurbuchen}, T.~H. 2008,
  \href{http://dx.doi.org/10.1086/589738}{\JournalTitle{\apj}, 683, 1192}

\bibitem[{{Lynch} {et~al.}(2009){Lynch}, {Antiochos}, {Li}, {Luhmann}, \&
  {DeVore}}]{lynch09}
{Lynch}, B.~J., {Antiochos}, S.~K., {Li}, Y., {Luhmann}, J.~G., \& {DeVore},
  C.~R. 2009,
  \href{http://dx.doi.org/10.1088/0004-637X/697/2/1918}{\JournalTitle{\apj},
  697, 1918}

\bibitem[{{Ma} {et~al.}(2011){Ma}, {Raymond}, {Golub}, {Lin}, {Chen}, {Grigis},
  {Testa}, \& {Long}}]{ma11}
{Ma}, S., {Raymond}, J.~C., {Golub}, L., {et~al.} 2011,
  \href{http://dx.doi.org/10.1088/0004-637X/738/2/160}{\JournalTitle{\apj},
  738, 160}

\bibitem[{{Manchester} {et~al.}(2017){Manchester}, {Kilpua}, {Liu}, {Lugaz},
  {Riley}, {T{\"o}r{\"o}k}, \& {Vr{\v{s}}nak}}]{manchester17}
{Manchester}, W., {Kilpua}, E. K.~J., {Liu}, Y.~D., {et~al.} 2017,
  \href{http://dx.doi.org/10.1007/s11214-017-0394-0}{\JournalTitle{\ssr}, 212,
  1159}

\bibitem[{{Mays} {et~al.}(2015){Mays}, {Thompson}, {Jian}, {Colaninno},
  {Odstrcil}, {M{\"o}stl}, {Temmer}, {Savani}, {Collinson}, {Taktakishvili},
  {MacNeice}, \& {Zheng}}]{mays15}
{Mays}, M.~L., {Thompson}, B.~J., {Jian}, L.~K., {et~al.} 2015,
  \href{http://dx.doi.org/10.1088/0004-637X/812/2/145}{\JournalTitle{\apj},
  812, 145}

\bibitem[{{Miki{\'c}} \& {Linker}(1996)}]{mikic94}
{Miki{\'c}}, Z., \& {Linker}, J.~A. 1996,
  \href{http://dx.doi.org/10.1063/1.51370}{in American Institute of Physics
  Conference Series, Vol. 382, Proceedings of the eigth International solar
  wind Conference: Solar wind eight, ed. D.~{Winterhalter}, J.~T. {Gosling},
  S.~R. {Habbal}, W.~S. {Kurth}, \& M.~{Neugebauer}}, 104

\bibitem[{{Mohamed} {et~al.}(2012){Mohamed}, {Gopalswamy}, {Yashiro},
  {Akiyama}, {M{\"a}kel{\"a}}, {Xie}, \& {Jung}}]{mohamed12}
{Mohamed}, A.~A., {Gopalswamy}, N., {Yashiro}, S., {et~al.} 2012,
  \href{http://dx.doi.org/10.1029/2011JA016589}{\JournalTitle{\jgra}, 117,
  A01103}

\bibitem[{{Moore} {et~al.}(2001){Moore}, {Sterling}, {Hudson}, \&
  {Lemen}}]{moore01}
{Moore}, R.~L., {Sterling}, A.~C., {Hudson}, H.~S., \& {Lemen}, J.~R. 2001,
  \href{http://dx.doi.org/10.1086/320559}{\JournalTitle{\apj}, 552, 833}

\bibitem[{{M{\"o}stl} {et~al.}(2015){M{\"o}stl}, {Rollett}, {Frahm}, {Liu},
  {Long}, {Colaninno}, {Reiss}, {Temmer}, {Farrugia}, {Posner}, {Dumbovi{\'c}},
  {Janvier}, {D{\'e}moulin}, {Boakes}, {Devos}, {Kraaikamp}, {Mays}, \&
  {Vr{\v{s}}nak}}]{moestl15}
{M{\"o}stl}, C., {Rollett}, T., {Frahm}, R.~A., {et~al.} 2015,
  \href{http://dx.doi.org/10.1038/ncomms8135}{\JournalTitle{\natco}, 6, 7135}

\bibitem[{{Panasenco} {et~al.}(2013){Panasenco}, {Martin}, {Velli}, \&
  {Vourlidas}}]{panasenco13}
{Panasenco}, O., {Martin}, S.~F., {Velli}, M., \& {Vourlidas}, A. 2013,
  \href{http://dx.doi.org/10.1007/s11207-012-0194-3}{\JournalTitle{\solphys},
  287, 391}

\bibitem[{{Posner} {et~al.}(2021){Posner}, {Arge}, {Staub}, {StCyr}, {Folta},
  {Solanki}, {Strauss}, {Effenberger}, {Gandorfer}, {Heber}, {Henney},
  {Hirzberger}, {Jones}, {K{\"u}hl}, {Malandraki}, \& {Sterken}}]{posner21}
{Posner}, A., {Arge}, C.~N., {Staub}, J., {et~al.} 2021,
  \href{http://dx.doi.org/10.1029/2021SW002777}{\JournalTitle{\spwea}, 19,
  e2021SW002777}

\bibitem[{{Priest} \& {D{\'e}moulin}(1995)}]{priest95}
{Priest}, E.~R., \& {D{\'e}moulin}, P. 1995,
  \href{http://dx.doi.org/10.1029/95JA02740}{\JournalTitle{\jgr}, 100, 23443}

\bibitem[{{Pulkkinen}(2007)}]{pulkkinen07}
{Pulkkinen}, T. 2007,
  \href{http://dx.doi.org/10.12942/lrsp-2007-1}{\JournalTitle{\lrsp}, 4, 1}

\bibitem[{{Reinard} \& {Fisk}(2004)}]{reinard04}
{Reinard}, A.~A., \& {Fisk}, L.~A. 2004,
  \href{http://dx.doi.org/10.1086/392493}{\JournalTitle{\apj}, 608, 533}

\bibitem[{{Sahade} {et~al.}(2021){Sahade}, {C{\'e}cere}, {Costa}, \&
  {Cremades}}]{sahade21}
{Sahade}, A., {C{\'e}cere}, M., {Costa}, A., \& {Cremades}, H. 2021,
  \href{http://dx.doi.org/10.1051/0004-6361/202141085}{\JournalTitle{\aap},
  652, A111}

\bibitem[{{Sahade} {et~al.}(2020){Sahade}, {C{\'e}cere}, \&
  {Krause}}]{sahade20}
{Sahade}, A., {C{\'e}cere}, M., \& {Krause}, G. 2020,
  \href{http://dx.doi.org/10.3847/1538-4357/ab8f25}{\JournalTitle{\apj}, 896,
  53}

\bibitem[{{Sahade} {et~al.}(2022){Sahade}, {C{\'e}cere}, {Sieyra}, {Krause},
  {Cremades}, \& {Costa}}]{sahade22}
{Sahade}, A., {C{\'e}cere}, M., {Sieyra}, M.~V., {et~al.} 2022,
  \href{http://dx.doi.org/10.1051/0004-6361/202243618}{\JournalTitle{\aap},
  662, A113}

\bibitem[{{Sahade} {et~al.}(2023){Sahade}, {Vourlidas}, {Balmaceda}, \&
  {C{\'e}cere}}]{sahade23}
{Sahade}, A., {Vourlidas}, A., {Balmaceda}, L.~A., \& {C{\'e}cere}, M. 2023,
  \href{http://dx.doi.org/10.3847/1538-4357/ace420}{\JournalTitle{\apj}, 953,
  150}

\bibitem[{{Shen} {et~al.}(2011){Shen}, {Wang}, {Gui}, {Ye}, \&
  {Wang}}]{shen.c11}
{Shen}, C., {Wang}, Y., {Gui}, B., {Ye}, P., \& {Wang}, S. 2011,
  \href{http://dx.doi.org/10.1007/s11207-011-9715-8}{\JournalTitle{\solphys},
  269, 389}

\bibitem[{{Sieyra} {et~al.}(2020){Sieyra}, {C{\'e}cere}, {Cremades},
  {Iglesias}, {Sahade}, {Mierla}, {Stenborg}, {Costa}, {West}, \&
  {D'Huys}}]{sierya20}
{Sieyra}, M.~V., {C{\'e}cere}, M., {Cremades}, H., {et~al.} 2020,
  \href{http://dx.doi.org/10.1007/s11207-020-01694-0}{\JournalTitle{\solphys},
  295, 126}

\bibitem[{{Siscoe} \& {Schwenn}(2006)}]{siscoe06}
{Siscoe}, G., \& {Schwenn}, R. 2006,
  \href{http://dx.doi.org/10.1007/s11214-006-9024-y}{\JournalTitle{\ssr}, 123,
  453}

\bibitem[{{Srivastava} \& {Venkatakrishnan}(2004)}]{srivastava04}
{Srivastava}, N., \& {Venkatakrishnan}, P. 2004,
  \href{http://dx.doi.org/10.1029/2003JA010175}{\JournalTitle{\jgr}, 109,
  A10103}

\bibitem[{{Talpeanu} {et~al.}(2022){Talpeanu}, {Poedts}, {D'Huys}, {Mierla}, \&
  {Richardson}}]{talpeanu22}
{Talpeanu}, D.~C., {Poedts}, S., {D'Huys}, E., {Mierla}, M., \& {Richardson},
  I.~G. 2022,
  \href{http://dx.doi.org/10.1051/0004-6361/202243150}{\JournalTitle{\aap},
  663, A32}

\bibitem[{{Temmer}(2021)}]{temmer21}
{Temmer}, M. 2021,
  \href{http://dx.doi.org/10.1007/s41116-021-00030-3}{\JournalTitle{\lrsp}, 18,
  4}

\bibitem[{{Thernisien}(2011)}]{thernisien11}
{Thernisien}, A. 2011,
  \href{http://dx.doi.org/10.1088/0067-0049/194/2/33}{\JournalTitle{\apjs},
  194, 33}

\bibitem[{{Thernisien} {et~al.}(2009){Thernisien}, {Vourlidas}, \&
  {Howard}}]{thernisien09}
{Thernisien}, A., {Vourlidas}, A., \& {Howard}, R.~A. 2009,
  \href{http://dx.doi.org/10.1007/s11207-009-9346-5}{\JournalTitle{\solphys},
  256, 111}

\bibitem[{{Titov} \& {D{\'e}moulin}(1999)}]{titov99}
{Titov}, V.~S., \& {D{\'e}moulin}, P. 1999, \JournalTitle{\aap}, 351, 707

\bibitem[{{Titov} {et~al.}(2022){Titov}, {Downs}, {T{\"o}r{\"o}k}, \&
  {Linker}}]{titov22}
{Titov}, V.~S., {Downs}, C., {T{\"o}r{\"o}k}, T., \& {Linker}, J.~A. 2022,
  \href{http://dx.doi.org/10.3847/1538-4357/ac874e}{\JournalTitle{\apj}, 936,
  121}

\bibitem[{{Titov} {et~al.}(2014){Titov}, {T{\"o}r{\"o}k}, {Mikic}, \&
  {Linker}}]{titov14}
{Titov}, V.~S., {T{\"o}r{\"o}k}, T., {Mikic}, Z., \& {Linker}, J.~A. 2014,
  \href{http://dx.doi.org/10.1088/0004-637X/790/2/163}{\JournalTitle{\apj},
  790, 163}

\bibitem[{{T{\"o}r{\"o}k} {et~al.}(2010){T{\"o}r{\"o}k}, {Berger}, \&
  {Kliem}}]{torok10}
{T{\"o}r{\"o}k}, T., {Berger}, M.~A., \& {Kliem}, B. 2010,
  \href{http://dx.doi.org/10.1051/0004-6361/200913578}{\JournalTitle{\aap},
  516, A49}

\bibitem[{{T{\"o}r{\"o}k} \& {Kliem}(2003)}]{torok03}
{T{\"o}r{\"o}k}, T., \& {Kliem}, B. 2003,
  \href{http://dx.doi.org/10.1051/0004-6361:20030692}{\JournalTitle{\aap}, 406,
  1043}

\bibitem[{{T{\"o}r{\"o}k} {et~al.}(2011){T{\"o}r{\"o}k}, {Panasenco}, {Titov},
  {Miki{\'c}}, {Reeves}, {Velli}, {Linker}, \& {De Toma}}]{torok11}
{T{\"o}r{\"o}k}, T., {Panasenco}, O., {Titov}, V.~S., {et~al.} 2011,
  \href{http://dx.doi.org/10.1088/2041-8205/739/2/L63}{\JournalTitle{\apjl},
  739, L63}

\bibitem[{{T{\"o}r{\"o}k} {et~al.}(2013){T{\"o}r{\"o}k}, {Temmer}, {Valori},
  {Veronig}, {van Driel-Gesztelyi}, \& {Vr{\v{s}}nak}}]{torok13}
{T{\"o}r{\"o}k}, T., {Temmer}, M., {Valori}, G., {et~al.} 2013,
  \href{http://dx.doi.org/10.1007/s11207-013-0269-9}{\JournalTitle{\solphys},
  286, 453}

\bibitem[{{T{\"o}r{\"o}k} {et~al.}(2018){T{\"o}r{\"o}k}, {Downs}, {Linker},
  {Lionello}, {Titov}, {Miki{\'c}}, {Riley}, {Caplan}, \& {Wijaya}}]{torok18}
{T{\"o}r{\"o}k}, T., {Downs}, C., {Linker}, J.~A., {et~al.} 2018,
  \href{http://dx.doi.org/10.3847/1538-4357/aab36d}{\JournalTitle{\apj}, 856,
  75}

\bibitem[{{Verbeke} {et~al.}(2019){Verbeke}, {Mays}, {Temmer}, {Bingham},
  {Steenburgh}, {Dumbovi{\'c}}, {N{\'u}{\~n}ez}, {Jian}, {Hess}, {Wiegand},
  {Taktakishvili}, \& {Andries}}]{verbeke19}
{Verbeke}, C., {Mays}, M.~L., {Temmer}, M., {et~al.} 2019,
  \href{http://dx.doi.org/10.1029/2018SW002046}{\JournalTitle{\spwea}, 17, 6}

\bibitem[{Verbeke {et~al.}(2022)Verbeke, {Mays}, Kay, Riley, Palmerio,
  Dumbović, Mierla, Scolini, Temmer, Paouris, Balmaceda, Cremades, \&
  Hinterreiter}]{verbeke22}
Verbeke, C., {Mays}, M.~L., Kay, C., {et~al.} 2022,
  \href{http://dx.doi.org/https://doi.org/10.1016/j.asr.2022.08.056}{\JournalTitle{\adv},
  in press}

\bibitem[{{Vourlidas}(2015)}]{vourlidas15}
{Vourlidas}, A. 2015,
  \href{http://dx.doi.org/10.1002/2015SW001173}{\JournalTitle{\spwea}, 13, 197}

\bibitem[{{Vr{\v{s}}nak}(2001)}]{vrsnak01}
{Vr{\v{s}}nak}, B. 2001,
  \href{http://dx.doi.org/10.1029/2000JA004007}{\JournalTitle{\jgr}, 106,
  25249}

\bibitem[{{Wang} {et~al.}(2020){Wang}, {Hoeksema}, \& {Liu}}]{wang.j.20a}
{Wang}, J., {Hoeksema}, J.~T., \& {Liu}, S. 2020,
  \href{http://dx.doi.org/10.1029/2019JA027530}{\JournalTitle{\jgra}, 125,
  e2019JA027530}

\bibitem[{{Wang} {et~al.}(2022){Wang}, {Liu}, \& {Luo}}]{wang.j.22a}
{Wang}, J., {Liu}, S., \& {Luo}, B. 2022,
  \href{http://dx.doi.org/https://doi.org/10.1016/j.asr.2022.06.017}{\JournalTitle{\adv},
  in press}

\bibitem[{{Wang} {et~al.}(2015){Wang}, {Liu}, {Dai}, {Yang}, {Huang}, \&
  {Hu}}]{wang.r15}
{Wang}, R., {Liu}, Y.~D., {Dai}, X., {et~al.} 2015,
  \href{http://dx.doi.org/10.1088/0004-637X/814/1/80}{\JournalTitle{\apj}, 814,
  80}

\bibitem[{{Wang} {et~al.}(2004){Wang}, {Shen}, {Wang}, \& {Ye}}]{wang.y04}
{Wang}, Y., {Shen}, C., {Wang}, S., \& {Ye}, P. 2004,
  \href{http://dx.doi.org/10.1023/B:SOLA.0000043576.21942.aa}{\JournalTitle{\solphys},
  222, 329}

\bibitem[{{Wold} {et~al.}(2018){Wold}, {Mays}, {Taktakishvili}, {Jian},
  {Odstrcil}, \& {MacNeice}}]{wold18}
{Wold}, A.~M., {Mays}, M.~L., {Taktakishvili}, A., {et~al.} 2018,
  \href{http://dx.doi.org/10.1051/swsc/2018005}{\JournalTitle{\jswsc}, 8, A17}

\bibitem[{{Xie} {et~al.}(2009){Xie}, {St.~Cyr}, {Gopalswamy}, {Yashiro},
  {Krall}, {Kramar}, \& {Davila}}]{xie09}
{Xie}, H., {St.~Cyr}, O.~C., {Gopalswamy}, N., {et~al.} 2009,
  \href{http://dx.doi.org/10.1007/s11207-009-9422-x}{\JournalTitle{\solphys},
  259, 143}

\bibitem[{{Xiong} {et~al.}(2009){Xiong}, {Zheng}, \& {Wang}}]{xiong09}
{Xiong}, M., {Zheng}, H., \& {Wang}, S. 2009,
  \href{http://dx.doi.org/10.1029/2009JA014079}{\JournalTitle{\jgr}, 114,
  A11101}

\bibitem[{{Yeh}(1983)}]{yeh83}
{Yeh}, T. 1983, \href{http://dx.doi.org/10.1086/160635}{\JournalTitle{\apj},
  264, 630}

\bibitem[{{Zhang} {et~al.}(2021){Zhang}, {Temmer}, {Gopalswamy}, {Malandraki},
  {Nitta}, {Patsourakos}, {Shen}, {Vr{\v{s}}nak}, {Wang}, {Webb}, {Desai},
  {Dissauer}, {Dresing}, {Dumbovi{\'c}}, {Feng}, {Heinemann}, {Laurenza},
  {Lugaz}, \& {Zhuang}}]{zhang.j21}
{Zhang}, J., {Temmer}, M., {Gopalswamy}, N., {et~al.} 2021,
  \href{http://dx.doi.org/10.1186/s40645-021-00426-7}{\JournalTitle{\peps}, 8,
  56}

\bibitem[{{Zhou} {et~al.}(2023){Zhou}, {Jiang}, {Yu}, {Wang}, {Hao}, \&
  {Cui}}]{zhou23}
{Zhou}, Z., {Jiang}, C., {Yu}, X., {et~al.} 2023,
  \href{http://dx.doi.org/10.3389/fphy.2023.1119637}{\JournalTitle{\frp}, 11,
  1119637}

\bibitem[{{Zhuang} {et~al.}(2019){Zhuang}, {Wang}, {Hu}, {Shen}, {Liu}, {Gou},
  {Zhang}, \& {Li}}]{zhuang19}
{Zhuang}, B., {Wang}, Y., {Hu}, Y., {et~al.} 2019,
  \href{http://dx.doi.org/10.3847/1538-4357/ab139e}{\JournalTitle{\apj}, 876,
  73}

\bibitem[{{Zuccarello} {et~al.}(2012){Zuccarello}, {Bemporad}, {Jacobs},
  {Mierla}, {Poedts}, \& {Zuccarello}}]{zuccarello12}
{Zuccarello}, F.~P., {Bemporad}, A., {Jacobs}, C., {et~al.} 2012,
  \href{http://dx.doi.org/10.1088/0004-637X/744/1/66}{\JournalTitle{\apj}, 744,
  66}

\end{thebibliography}

\end{document}